\documentclass[aps,nofootinbib,pra,notitlepage,twocolumn,superscriptaddress]{revtex4-1}
\usepackage{amsfonts,amsmath,amssymb,amsthm}
\usepackage{graphicx}
\usepackage{xspace}
\usepackage[table]{xcolor}
\usepackage{tabularx}

\newcommand{\rrangle}{\rangle\!\rangle} \newcommand{\llangle}{\langle\!\langle}

\newcommand{\complex}{\mathbb{C}}
\newcommand{\expect}[1]{\ensuremath{\left\langle#1\right\rangle}}
\newcommand{\ket}[1]{\ensuremath{\left|#1\right\rangle}}
\newcommand{\bra}[1]{\ensuremath{\left\langle#1\right|}}

\newcommand{\ketbra}[2]{\ket{#1}\!\!\bra{#2}}
\newcommand{\braopket}[3]{\ensuremath{\bra{#1}#2\ket{#3}}}
\newcommand{\proj}[1]{\ketbra{#1}{#1}}
\newcommand{\sket}[1]{\ensuremath{\left|#1\right\rrangle}}
\newcommand{\sbra}[1]{\ensuremath{\left\llangle#1\right|}}
\newcommand{\sbraket}[2]{\ensuremath{\left\llangle#1|#2\right\rrangle}}

\newcommand{\sbraopket}[3]{\ensuremath{\sbra{#1}#2\sket{#3}}}

\def\Id{1\!\mathrm{l}}
\newcommand{\Tr}{\mathrm{Tr}}

\renewcommand{\bar}[1]{\overline{#1}}

\newcommand{\cP}{\mathcal{P}}

\newcommand{\cE}{\mathcal{E}}
\newcommand{\cL}{L}
\newcommand{\cU}{\mathcal{U}}
\newcommand{\cH}{\mathcal{H}}

\newcommand{\bL}{\mathbb{L}}
\newcommand{\bH}{\mathbb{H}}
\newcommand{\bC}{\mathbb{C}}
\newcommand{\bA}{\mathbb{A}}
\newcommand{\bS}{\mathbb{S}}
\newcommand{\bM}{\mathbb{M}}
\newcommand{\bQ}{\mathbb{Q}}

\newcommand{\Lind}{\cL_+}
\newcommand{\PJ}{\epsilon_J}
\mathchardef\mdash="2D

\begin{document}
\title{A Taxonomy of Small Markovian Errors}

\author{Robin Blume-Kohout}
\affiliation{Quantum Performance Laboratory, Sandia National Laboratories, Albuquerque, NM 87185 and Livermore, CA 94550}
\author{Marcus P. da Silva}
\affiliation{Microsoft Quantum, One Microsoft Way, Redmond, WA 98052}
\author{Erik Nielsen}
\affiliation{Quantum Performance Laboratory, Sandia National Laboratories, Albuquerque, NM 87185 and Livermore, CA 94550}
\author{Timothy Proctor}
\affiliation{Quantum Performance Laboratory, Sandia National Laboratories, Albuquerque, NM 87185 and Livermore, CA 94550}
\author{Kenneth Rudinger}
\affiliation{Quantum Performance Laboratory, Sandia National Laboratories, Albuquerque, NM 87185 and Livermore, CA 94550}
\author{Mohan Sarovar}
\affiliation{Sandia National Laboratories, Livermore, CA 94550}
\author{Kevin Young}
\affiliation{Quantum Performance Laboratory, Sandia National Laboratories, Albuquerque, NM 87185 and Livermore, CA 94550}

\date{\today}

\begin{abstract}
\noindent Errors in quantum logic gates are usually modeled by quantum process matrices (CPTP maps).  But process matrices can be opaque, and unwieldy.  We show how to transform a gate's process matrix into an error generator that represents the same information more usefully.  We construct a basis of simple and physically intuitive elementary error generators, classify them, and show how to represent any gate's error generator as a mixture of elementary error generators with various rates.  Finally, we show how to build a large variety of reduced models for gate errors by combining elementary error generators and/or entire subsectors of generator space.  We conclude with a few examples of reduced models, including one with just $9N^2$ parameters that describes almost all commonly predicted errors on an $N$-qubit processor.
\end{abstract}

\maketitle

An ideal quantum computation is implemented by a sequence of unitary operations -- quantum \emph{logic gates} -- applied to a register of qubits.  But the real quantum processors being built in experimental labs today are not ideal.  Their logic gates are imperfect.  Models of imperfect gates are used to predict the results of running computations \cite{Reiner2018-it, Willsch2017-oq, Proctor2020-cq, Georgopoulos2021-hs, Rines2019-er}, to measure progress toward specific goals like fault tolerant error correction \cite{Elder2020-ue, Andersen2019-lq, Bermudez2019-po, Gong2019-wf, Bultink2019-cx, Negnevitsky2018-ab, Trout2018-fy, Takita2017-wp, Takita2016-po, Corcoles2015-mt, Chow2014-bc, Barends2014-ap, Cory1998-qa, Wright2019-mx, Bermudez2017-vg, Egan2020-yo, Magesan2013-ep, Tomita2014-io, Gutierrez2013-ey,Pal2020-zx, Chen2021-ou}, and to understand how errors in gates may be reduced \cite{Blume-Kohout2017-kn} or mitigated \cite{Murali2019-bs, Bultrini2020-ro, Hu2020-si, Endo2018-kb, Song2019-fg, Tannu2018-bx, Murphy2019-lw}.  The standard model for an imperfect quantum logic gate is a \emph{quantum process matrix} specifying a completely positive, trace preserving (CPTP) map \cite{Chuang1997-vf}.  Under certain assumptions (see below), a process matrix provides a complete description of how the qubits' state space evolves when the gate is applied.  But in this role, quantum processes have two critical flaws.  They are not easy to interpret, and their complexity grows exponentially with the register's size.  This article constructs a representation that helps address these issues.

\begin{figure}[t!]
\includegraphics[width=1\columnwidth]{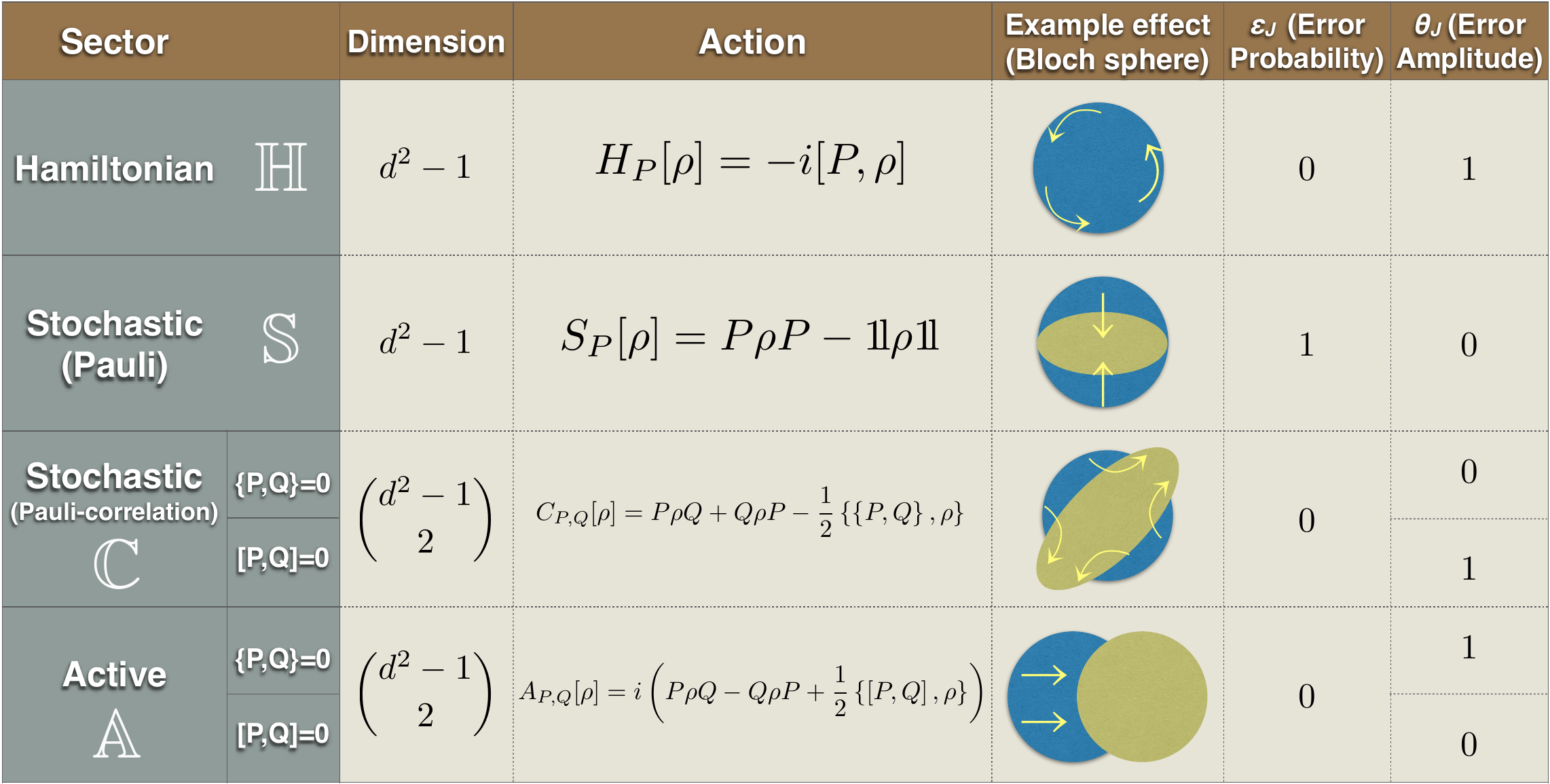}
\caption{We represent an imperfect gate by its \emph{error generator} (Sec.~\ref{sec:Generators}) $\cL = \log( G\bar{G}^{-1} )$, where $G$ is the process matrix describing the imperfect gate and $\bar{G}$ is the process matrix for the ideal gate.  We construct (Sec.~\ref{sec:Taxonomy}) a useful basis of \emph{elementary error generators} for the vector space $\bL$ containing $\cL$.  This basis defines a taxonomy of small Markovian errors, dividing generators into four sectors (subspaces) shown in this table along with their dimension and their Choi sum representation.  For each sector, we consider the single-qubit case (Sec.~\ref{sec:OneQubit}) and illustrate how the error process generated by a single element from that sector transforms the Bloch sphere.  Error metrics that quantify each elementary generator's production of incoherent and coherent errors (Sec. \ref{sec:Metrics}) are tabulated for the generators in each sector, and (for $C_{P,Q}$ and $A_{P,Q}$ generators) for the sub-cases where $P$ and $Q$ anticommute or commute.}
\label{fig:Table-v2}
\end{figure}

Our goal is to analyze and understand small, Markovian gate errors (Sec. \ref{sec:SmallMarkovian}).  We begin by representing such errors using \emph{error generators} (Sec. \ref{sec:Generators}).  Then, we classify all the error generators that describe small Markovian $N$-qubit errors (Sec. \ref{sec:Taxonomy}).  We show that any error generator can be written as a combination of three classes of elementary errors -- \emph{Hamiltonian}, \emph{stochastic}, and \emph{active} -- that are are invariant under unitary changes of basis.  We further divide the stochastic error generators into \emph{Pauli stochastic} and \emph{Pauli correlation} sectors, which are invariant under Clifford (but not arbitrary) unitaries.  We construct a complete basis of ``elementary error generators'' for each sector, using 1- and 2-qubit constructions as constructive examples.  We explain the physical origin and impact of each kind of error (Sec. \ref{sec:Taxonomy-Discussion}), and discuss the relationship between our error generators and the generators of Lindblad master equations (Sec. \ref{sec:Lindblad}).  After introducing simple metrics of coherent and incoherent error and tabulating them for each elementary error generator (Sec. \ref{sec:Metrics}), we show how to further partition those four main sectors into subsectors of fixed \emph{weight} and \emph{support} (Sec. \ref{sec:Reduced}).  This fine-grained partition of error generators into physically and logically meaningful classes, is the taxonomy promised in the title.  We conclude with what we see as the most exciting application of this framework:  the construction of customizable, efficient reduced models of errors in $N$-qubit logic operations that can describe and model specific errors or classes of errors in a quantum processor while minimizing the amount of resources wasted on unlikely or physically implausible errors.

\section{Small Markovian errors} \label{sec:SmallMarkovian}

We are interested in errors that are (1) small and (2) Markovian.  We begin by stating exactly what we mean by these terms.  Both represent idealized assumptions that never hold exactly in experiments, but can be tested experimentally, and are often approximately true.  Theorems and representations derived in the limit of small Markovian errors can provide accurate approximate results in real-world situations.

\subsection{Definitions}

We call a process that changes a system's state $\rho\to\rho'$ \emph{Markovian} if, given the nature of the process, $\rho'$ is completely determined by $\rho$.  So if the error associated with a particular gate $g$ is Markovian, then it is described by some map $G_g: \rho \to \rho'$ that does not depend on the time of day, other gates performed previously, or any other ``context'' variable.  It then follows from the rules of quantum theory that $G_g$ must be linear, completely positive, and trace-preserving -- and thus that $G_g$ can be represented by a process matrix.  Other definitions of ``Markovian'' appear in the extensive literature on Markovian and non-Markovian quantum dynamics (see, e.g., \cite{Breuer2016-th} and references therein), but the definition used here is common, and we state it explicitly to avoid confusion.

We say the error in an implemented gate $g$ is \emph{small} if the process matrix $G_g$ that describes its action is close to the ``target'' process matrix $\bar{G}_g$ that describes an ideal, perfect implementation of $g$.  More precisely, we want $G_g - \bar{G}_g$ to be small, so that expressions that are $O\left([G_g - \bar{G}_g]^2\right)$ can be neglected.  It is sufficient that $\left\|G_g - \bar{G}_g\right\|_{\diamond} \ll 1$ (more about the diamond norm can be found in \cite{Aharonov1998-rc}).

\subsection{Superoperators}

Consider a logic gate on an $N$-qubit register.  The Hilbert space $\cH$ of the $N$ qubits has $d=2^N$ dimensions, and is isomorphic to $\complex^d$.  The ideal unitary target gate can be described and represented by a $d \times d$ unitary matrix $U$ that acts on states $\ket\psi\in\cH$.  But noisy evolutions require the richer state space of $d\times d$ density matrices $\rho$, in which pure states ($\rho = \proj{\psi}$) and unitary evolution ($\rho \to U\rho U^\dagger$) are special cases.  More generally, $\rho \to G[\rho]$, where $G$ is a \emph{completely positive, trace preserving, linear map} on operators (or ``CPTP map'' for short).  To represent and analyze this action, we can represent $\rho$ by a column vector $\sket{\rho}$ in the $d^2$-dimensional space $\mathcal{L}(\cH)$ of $d\times d$ matrices.  Equipped with the inner product 
\begin{equation}
\sbraket{A}{B} \equiv \Tr A^\dagger B,
\end{equation}
this is called \emph{Hilbert-Schmidt space}.  Now $G$ can be represented by a $d^2\times d^2$ matrix that acts by matrix multiplication on $\sket{\rho}$:
\begin{equation}
\sket{\rho} \to G\sket{\rho}.
\end{equation}
This representation of $G$ is associative -- the consecutive application of $G$ and then $H$ is described by $HG$ -- and has been called the \emph{superoperator}, \emph{transfer matrix}, or \emph{Liouville} representation.

$G$ can be represented in any orthonormal basis of matrices.  We use the $N$-qubit Pauli basis\footnote{We see no obstacle to constructing a version of our framework for systems of Hilbert space dimension $d\neq 2^N$, but many of the details would need to be done in different ways because we make heavy use of (1) the Pauli basis and (2) underlying locality assumptions intrinsic to $N$ qubits.} $\cP = \{P_1\ldots P_{d^2}\}$.  It comprises all $N$-fold tensor products of the single-qubit Pauli group $\{\Id,X,Y,Z\}$.  They are all Hermitian.  For every $N$, $P_1=\Id$, and all the rest are traceless.  The orthonormality relation is $\Tr PQ = d\delta_{P,Q}$ for all $P,Q\in\cP$.  Every pair of $N$-qubit Paulis either commutes ($[P,Q]=0$) or anticommutes ($\{P,Q\}=0$), and each Pauli except $\Id$ commutes with exactly half the other Paulis and anticommutes with the rest.

An $N$-qubit superoperator $G$ written in this basis is a $4^N\times 4^N$ matrix\footnote{An $N$-qubit density matrix $\rho$ written in the Pauli basis is a column vector $\sket{\rho}$ with elements $\sbraket{P}{\rho} = \Tr P\rho$.} with elements
\begin{equation}
    G_{P,Q} = \sbraopket{P}{G}{Q} = \Tr P G[Q].
\end{equation}
A CPTP superoperator must preserve Hermiticity. Since Paulis are Hermitian, $G$ is a real matrix with $16^N$ free parameters.  The complete positivity (CP) constraint defines a cone\footnote{A cone is a subset of a vector space that is closed under \emph{positive} linear combinations.}, but does not reduce the dimension, so a CP map still has $16^N$ mostly-free parameters.  A trace-preserving (TP) map satisfies $G^T[\Id] = \Id$, so its top row must be $[1,0,\ldots 0]$, and CPTP maps have $4^N(4^N-1)$ free parameters.  A map $G$ is called \emph{unital} if $G[\Id] = \Id$, and $G$ is unital iff its leftmost column is $[1,0,\ldots 0]^T$.

\section{Error generators} \label{sec:Generators}

When $G$ describes an imperfect gate, we are less interested in $G$ itself than in how it differs from its unitary target $\bar{G}$.  So we focus on the set of possible small deviations from $\bar{G}$.  This set is different for each $\bar{G}$, but we can remove the variation by modeling an imperfect gate as its ideal unitary followed by a \emph{post-gate error process},
\begin{equation}
G = \cE \bar{G}.
\end{equation}
Now $\cE \equiv G \bar{G}^{-1}$ faithfully represents the error in $G$, and is always close to the identity process $\Id$ when $G$ is close to $\bar{G}$.  This transformation was suggested and explored by Korotkov in Ref. \cite{Korotkov2013-bz}, and deployed experimentally by Rodionov \emph{et al} in Ref. \cite{Rodionov2014-mt}.  The framework we construct here takes advantage of some ideas that first appeared in those papers.

If $\cE=\Id$, then the gate is perfect.  Small errors correspond to small deviations from $\Id$.  To isolate that deviation, we can compute $\cE - \Id$, or $\log\cE$.  These expressions become identical in the limit $\cE\to\Id$, since
\begin{equation}
    \log X = (X - \Id) + O\left((X-\Id)^2\right),
\end{equation}
but there are subtle and interesting differences, which we revisit in Section \ref{sec:Lindblad}.  For now, we choose the logarithm.  $\cE$ is a real matrix. For small errors it has a real logarithm (see below) with an unambiguous principal branch.  We define the \emph{post-gate error generator} for $G$ as 
\begin{equation}
    \cL = \log(\cE) \ \ \Leftrightarrow\ \ G = e^{\cL} \bar{G}.
\end{equation}
$\cL$ is a faithful representation of the error in $G$. But unlike $G$ itself, its magnitude and nature directly represent the magnitude and kind of errors in $G$.  It generates errors after $G$ in the same sense that a Hamiltonian $H$ generates a unitary $U = e^{iH}$ \cite{Lindblad1976-qr}.

It would be equally valid (and completely equivalent) to write $G = \bar{G} \cE'$, and use pre-gate error processes and error generators $\cL' = \log( \cE' )$.  (A representation using \emph{both} pre- and post-gate error processes was proposed by Wallman \cite{Wallman2018-wy}).  We use post-gate error generators because it is slightly more intuitive to imagine the error process occurring after the gate, rather than before it.

But this highlights another possible choice:  we could compute a \emph{during-gate error generator} $\cL''$ by writing $G = \exp( \log \bar{G} + \cL'' )$.  This representation works equally well for some errors, and it is physically well-motivated if $G$ was implemented by a simple pulse.  But unlike the pre- and post-gate generators, it is not always a faithful representation of arbitrary errors.  Some noisy processes $G \approx \bar{G}$ have no during-gate error generator.  For an example, let $\bar{G}[\rho] = Z\rho Z$ perform a single-qubit $Z_\pi$ rotation, and $G = \cE_{pX}\bar{G}$, where $\cE_{pX}[\rho] = (1-p)\rho + pX\rho X$ causes a stochastic $X$ error with probability $p>0$.  Now, $\bar{G}$ has two $-1$ eigenvalues (corresponding to the Pauli $X$ and $Y$ matrices) that form a Jordan block of size 2, so it has a real logarithm.  But the stochastic $X$ error process breaks this symmetry in $G$, which has two \emph{distinct} negative real eigenvalues, and therefore has no real logarithm at all\footnote{There is physics behind this math, too!  Any stochastic $X$ or $Y$ error generated continuously during this $\pi$ gate gets ``twirled'' into a perfectly equal mixture of $X$ and $Y$ errors.  So there really is no during-gate generator that can produce a biased mixture of $X$ and $Y$ errors afterward.}.

The post-gate error generator avoids this problem. It is the logarithm of a real matrix close to $\Id$, so it has no negative eigenvalues, and a real $\cL$ always exists.  We emphasize, however, that the post-gate error generator is not intended to model the exact mechanism that generated errors in $G$.  In many systems, gates are implemented by complex pulses with strongly time-varying characteristics, and complicated error mechanisms.  The error generators we consider here are a mathematical representation of their \emph{effects}, not necessarily of their cause.

\begin{figure}[t!]
\includegraphics[width=1\columnwidth]{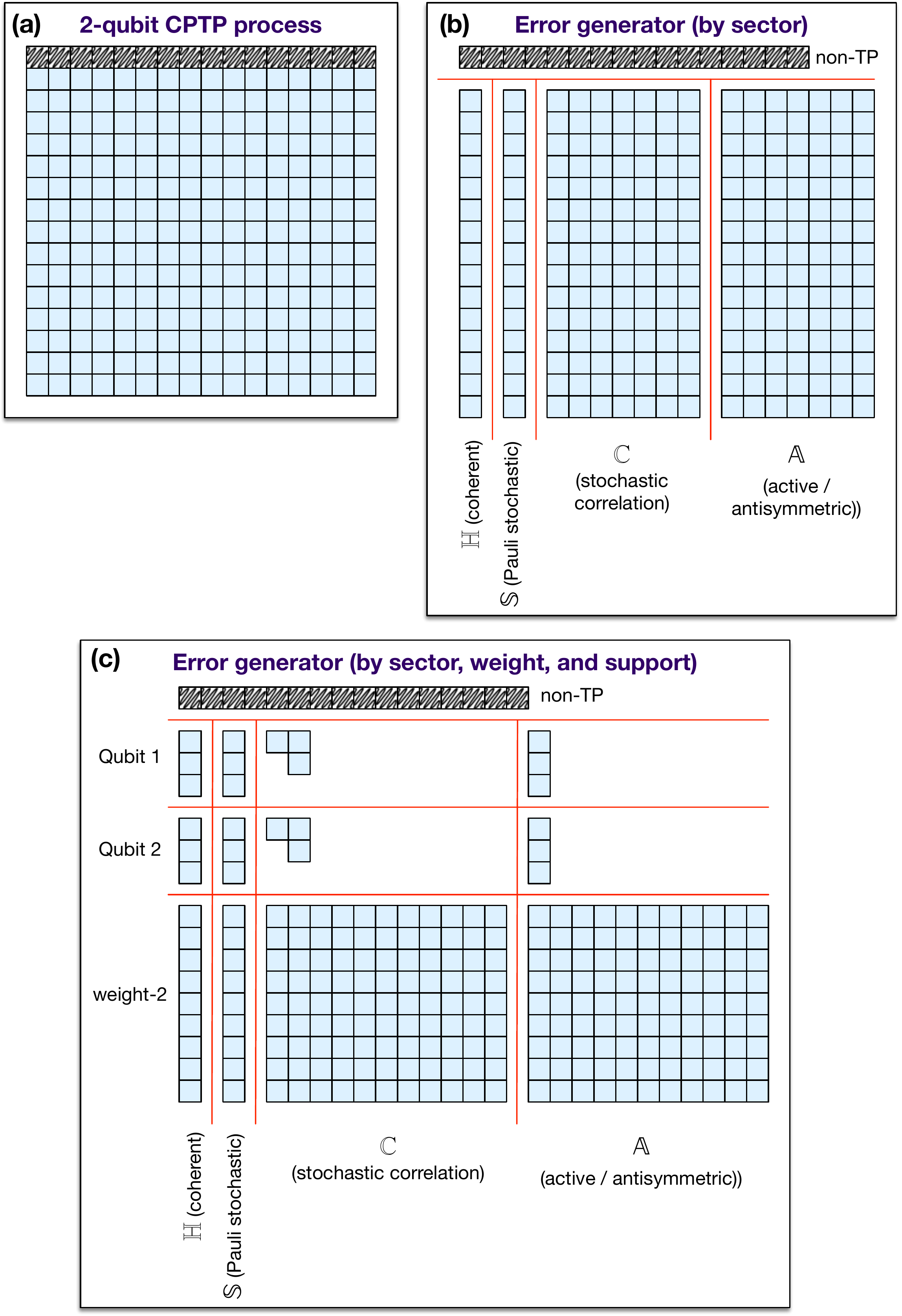}
\caption{A 2-qubit CPTP map \textbf{(a)} has 240 free parameters ($16\times 16$ minus $16$ for the top row, which is constrained by trace preservation).  It can be reparameterized by its error generator $\textbf{(b)}$, which can be split up into its projections onto $\mathbb{H}$, $\mathbb{S}$, $\mathbb{C}$, and $\mathbb{A}$ sectors.  Each of these sectors can be further partitioned (see Sec.~\ref{sec:Reduced}) into generators with a fixed \emph{weight} (number of qubits on which it acts) and \emph{support} (subset of qubits on which it acts).}
\label{fig:2Q-Partition}
\end{figure}

We work with generators instead of processes because switching representations changes the meaning of linear combination in a subtle and useful way.  Linear combination is natural for both processes and generators, but linear combinations of processes are \emph{convex} combinations, a.k.a.~mixtures.  $(\cE_1 + \cE_2)/2$ means ``Flip a coin and perform $\cE_1$ \emph{or} $\cE_2$.''  In contrast, linear combinations of generators indicate \emph{composition}.  If $H_1$ and $H_2$ are Hamiltonians, then $H_1$ + $H_2$ means ``Apply $H_1$ \emph{and} $H_2$,'' and it generates a unitary operation, not a mixture of unitaries.  Although composition and mixture converge in the small error regime, generator space admits a clean partition into subspaces representing distinct (and potentially concurrent) error mechanisms, in a way that the convex set of processes does not.

\section{A taxonomy of error generators} \label{sec:Taxonomy}

$\cL$ presents the same information as $G$, but is more amenable to analysis.  Error generators can be dissected into a list of easily interpretable terms.  Whereas quantum processes like $G$ form a \emph{semigroup} \cite{Lindblad1976-qr}, generators like $\cL$ form a Lie \emph{semialgebra}\footnote{A semialgebra is like an algebra, but restricted to a cone.} that is the solid tangent cone\footnote{The solid tangent cone to a convex set $X$, at a point $x$ on its boundary, is the closure of the cone formed by all rays from $x$ through a distinct point in $X$.} to the set of CPTP maps at $\Id$. Its linear closure is a $d^2(d^2-1)$-dimensional vector space that we call \emph{generator space} ($\bL$). 

We can construct a basis for $\bL$ in which each element has a simple interpretation and produces a recognized quantum logical error.  We call these \emph{elementary generators}.  They fall into four classes that define subspaces of $\bL$.  We denote these subspaces by $\bH$, $\bS$, $\bC$, and $\bA$ (see Fig.~\ref{fig:2Q-Partition}a-b).  The elementary generators in $\bH$ and $\bS$ are indexed by a Pauli operator $P$, and we denote them by $\{H_P\}$ and $\{S_P\}$ respectively.  Elementary generators in the other two classes are indexed by distinct pairs $(P,Q)$ of distinct Paulis\footnote{We denote the set of \emph{distinct} pairs of distinct Pauli operators as $\{(P,Q>P)\}$, to indicate that only one of $(P,Q)$ or $(Q,P)$ is included.}, and denoted by $\{C_{P,Q}\}$ and $\{A_{P,Q}\}$ respectively.  We can write any $\cL\in\bL$ as a linear combination of elementary generators with real coefficients,
\begin{eqnarray}
    \cL &=& \cL_{\mathbb{H}} + \cL_{\mathbb{S}} + \cL_{\mathbb{C}} + \cL_{\mathbb{A}} \label{eq:LHSCA}\\
    &=& \sum_{P}{h_P H_P} + \sum_P{s_P S_P} \nonumber \\
    &&+\sum_{P,Q>P}{c_{P,Q} C_{P,Q}} + \sum_{P,Q>P}{a_{P,Q} A_{P,Q}}. \label{eq:Lsum}
\end{eqnarray}
We refer to each coefficient as the \emph{rate} of the corresponding error process, in keeping with its appearance in the exponent of $\cE = e^{\cL}$.

\subsection{Choi sums and units}

The easiest way to define these elementary generators is using another commonly-used representation that we call the \emph{Choi sum} representation \cite{Johnston2011-nc, Choi1975-hx}:
\begin{equation}
G[\rho] = \sum_{P,Q\in\cP}{\chi_{P,Q} P \rho Q} \label{eq:Choi}.
\end{equation}
There is a $\chi$-matrix representation for every superoperator $G$.  Equation \ref{eq:Choi} can be seen as an expansion of $G$ in a complete orthogonal basis of superoperators that we call \emph{Choi units}, defined by $X_{P,Q}[\rho] = P \rho Q$.  It is easy to show that the Choi units are mutually orthogonal (by the Hilbert-Schmidt inner product defined on superoperators), and since there are $d^4 = 16^N$ of them, they form a complete basis.  The Choi sum representation can be defined with respect to any operator basis (a common choice is the basis of matrix units $\{\ketbra{i}{j}\}$), but we only use the Pauli basis here.  The best-known property of the Choi sum representation is that $G$ is CP iff $\chi\geq0$ \cite{Choi1975-hx}.

We now define elementary generators, in three steps.  

\subsection{Hamiltonian generators}

\begin{figure}[t!]
\includegraphics[width=0.49\columnwidth]{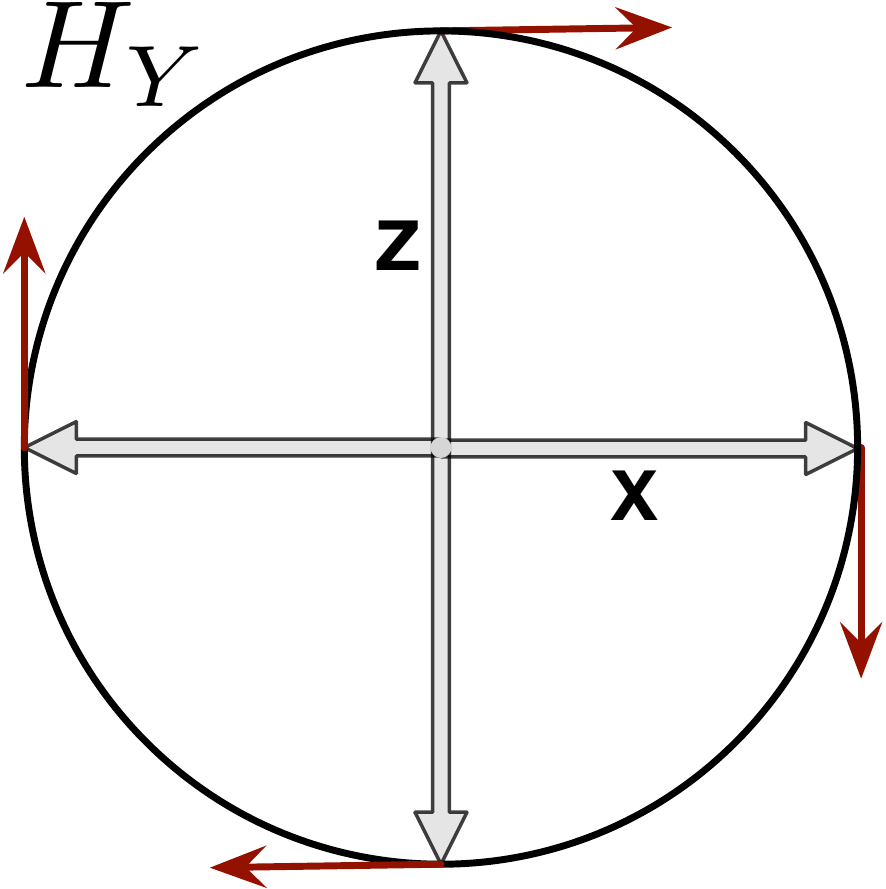}
\caption{The action of a representative Hamiltonian error generator.  This is how the 1-qubit $H_Y$ generator acts on the $X\mdash Z$ plane of the Bloch sphere.  It generates rotations, sending $Z\to X$ and $X\to -Z$.}
\label{fig:HAction}
\end{figure}

First we consider unitary error processes $\cE[\rho] = U\rho U^\dagger$.  Their generators are well-known; if $U = e^{-iJ}$ then $\cE[\rho] = e^{H_J}[\rho]$, where $H_J[\rho] = -i[J,\rho]$.  Since any Hamiltonian $J$ can be expanded in the Pauli basis as $J = \sum_P{h_P P}$, we can write $H_J = \sum_P{h_P H_P}$, and so the generator of any unitary error can be written as a linear combination of $d^2-1$ \emph{Hamiltonian generators}
\begin{equation}
    H_P[\rho] = -i[P,\rho] = -iP\rho \Id + i\Id\rho P,\label{eq:HP}
\end{equation}
where the last expression is explicitly a Choi sum.  $\bH$ is the $(d^2-1)$-dimensional subspace of $\bL$ spanned by the Hamiltonian generators.  It is invariant under unitary changes of basis -- i.e., if $\cE = e^{\sum_P{h_PH_P}}$, then for any unitary matrix $U$ whose superoperator representation is $\cU[\cdot] = U\cdot U^\dagger$, $\cU\cE\cU^\dagger = e^{\sum_P{h'_PH_P}}$ for some set of coefficients $\{h'_P\}$.

\subsection{Stochastic generators}

\begin{figure}[t!]
\includegraphics[width=1\columnwidth]{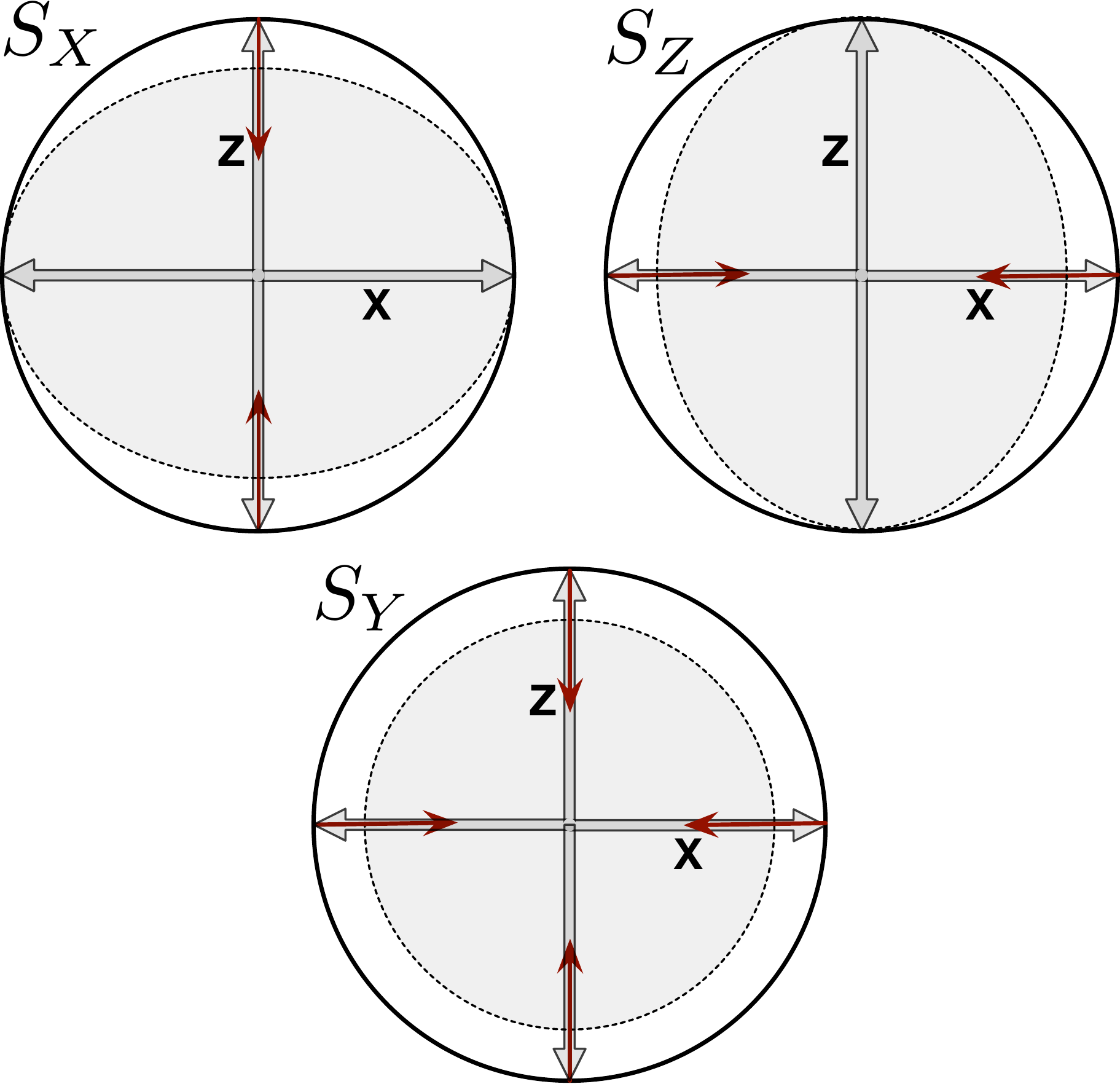}
\caption{The action of 1-qubit Pauli-stochastic error generators on the $X\mdash Z$ plane of the Bloch sphere.  $S_X$ shrinks $Z\to 0$ but leaves $X$ unchanged.  $S_Z$ shrinks $X\to 0$ but leaves $Z$ unchanged.  $S_Y$ shrinks both $X$ and $Z$.  Each generator, by itself, generates a dephasing process.  The sum of all three generates depolarization.}
\label{fig:SAction}
\end{figure}

Second, we consider the effect of convex mixtures of Hamiltonian generators.  This does not mean that the system is driven by a linear combination of Hamiltonians, $J = \sum_k{J_k}$.  Instead, the system's dynamical evolution is a linear combination of the evolutions resulting from those $J_k$, i.e.
\begin{equation}
\cE[\rho] = \sum_{k}{e^{-iJ_k}\rho e^{iJ_k}}.
\end{equation}
We treat the $\{J_k\}$ as small, and expand in powers of them.  To first order in $\{J_k\}$, the corresponding generator is indeed just $H_J = \sum_k{H_{J_k}}$.  But including second order terms yields
\begin{equation}
\cE[\rho] \approx \rho + H_J[\rho] + \sum_{k}{J_k\rho J_k} - \frac12\left\{J_k^2,\rho\right\}.
\end{equation}
Expanding each $J_k$ in the Pauli basis and rearranging yields a sum of the form
\begin{eqnarray}
\cE[\rho] \approx &\rho& + H_J[\rho] + \sum_{P}{s_P \left( P\rho P - \rho\right)} \\ &+& \sum_{P,Q}{c_{P,Q}\left(P \rho Q + Q\rho P - \frac12\left\{\{P,Q\},\rho\right\}\right)},\nonumber
\end{eqnarray}
for some $s_P$ and $c_{P,Q}$ coefficients that can be computed from the Pauli expansions of the $J_k$.  So \emph{any} convex mixture of unitary evolutions is generated by (1) the Hamiltonian generators discussed above, plus (2) some linear combination of $d^2-1$ \emph{stochastic Pauli generators} indexed by Pauli operators $P$,
\begin{equation}
S_P[\rho] = P\rho P - \Id\rho\Id,\label{eq:SP}
\end{equation}
and $(d^2-1)(d^2-2)/2$ \emph{Pauli correlation generators} indexed by distinct pairs of non-identity Paulis $(P,Q)$,
\begin{equation}
C_{P,Q}[\rho] = P\rho Q + Q\rho P - \frac12\left\{\left\{P,Q\right\},\rho\right\}.\label{eq:CPQ}
\end{equation}
The $\bS$ and $\bC$ subspaces are spanned by the $S$ and $C$ generators, respectively. It is easy to show that each of these generators can be independently varied, just by considering mixtures of $e^{iJ}$ and $e^{-iJ}$ with $J \propto P$ or $J \propto P\pm Q$.  The union of $\bS$ and $\bC$ is a $d^2(d^2-1)/2$-dimensional subspace of stochastic generators.  It is also invariant under unitary changes of basis.

\begin{figure}[t!]
\includegraphics[width=0.69\columnwidth]{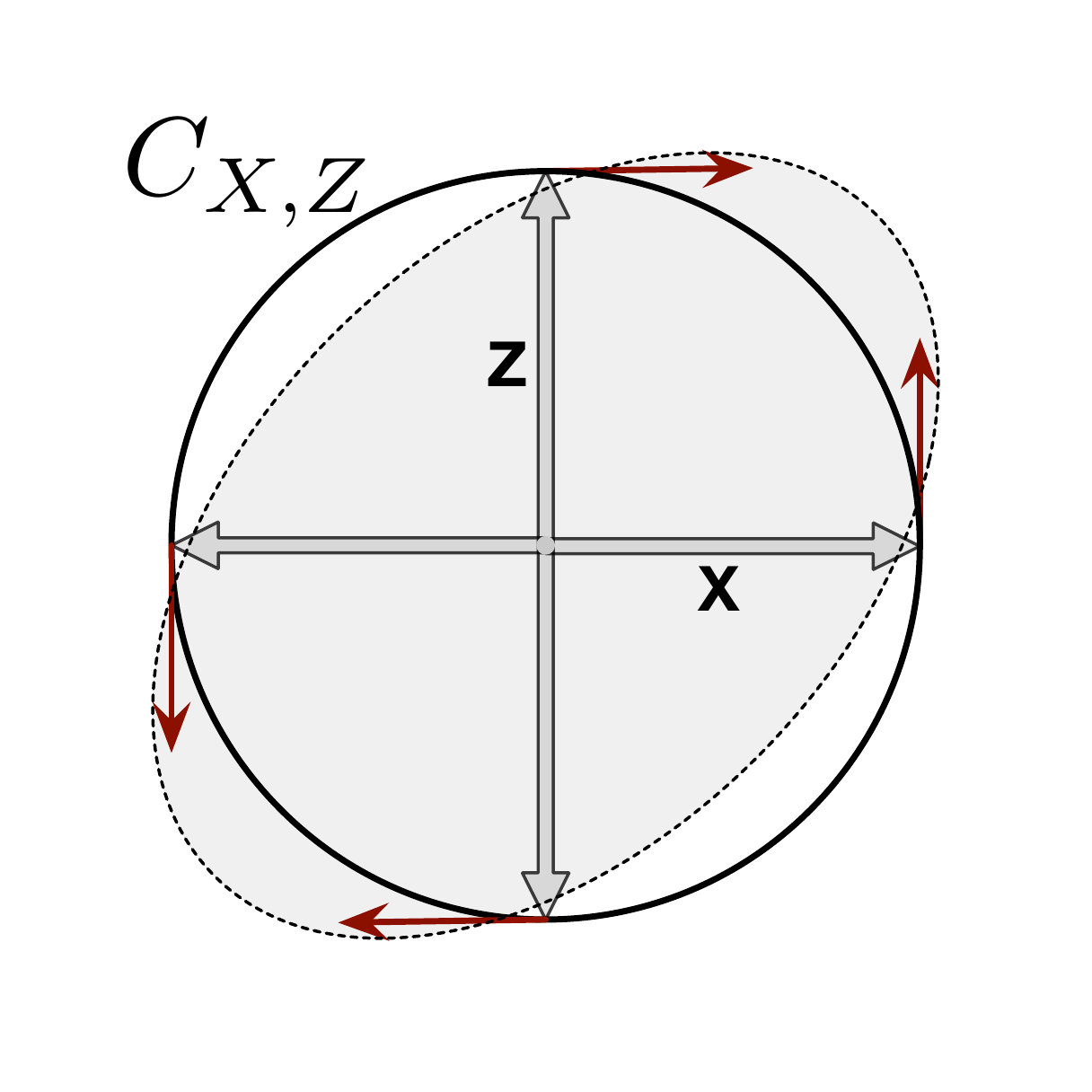}
\caption{The action of a representative Pauli-correlation error generator.  This is how the 1-qubit $C_{X,Z}$ generator acts on the $X\mdash Z$ plane of the Bloch sphere.  Its action on $Z$ is identical to $H_Y$, taking $Z\to X$, but instead of taking $X\to -Z$ it takes $X\to Z$.  The result is a squeezing transformation that stretches the $X+Z$ axis and shrinks the $X-Z$ axis.}
\label{fig:CAction}
\end{figure}

\subsection{Active (antisymmetric) generators}

Third, we consider everything that is left.  $\bL$ has $d^2(d^2-1)$ dimensions, and we have constructed disjoint subspaces $\bH$ ($d^2-1$ dimensions), $\bS$ ($d^2-1$ dimensions), and $\bC$ [$(d^2-1)(d^2-2)/2$ dimensions].  Their complement, then, has $(d^2-1)(d^2-2)/2$ dimensions.

To construct elementary generators for this subspace, we consider how the $H$, $S$, and $C$ elementary generators relate to the Choi units $X_{P,Q}$.  The Choi units span the entire $d^2$-dimensional space of superoperators, which is larger than $\bL$ because it contains non-TP processes.  Each of the stochastic $S$ and $C$ generators is a linear combination of \emph{symmetrized} Choi units -- i.e., $X_{P,P}$ or $X_{P,Q} + X_{Q,P}$.  The Hamiltonian $H$ generators are antisymmetrized linear combinations proportional to $X_{P,\Id} - X_{\Id,P}$.  All of $H$, $S$, and $C$ generators are orthogonal to all the antisymmetrized Choi units of the form $X_{P,Q} - X_{Q,P}$ with $P,Q \neq \Id$.  Informed by this observation, we construct $(d^2-1)(d^2-2)/2$ additional error generators indexed by distinct pairs of non-identity Paulis,
\begin{equation}
    A_{P,Q}[\rho] = i\left( P \rho Q -  Q\rho P + \frac{1}{2}\left\{\left[P,Q\right],\rho\right\}\right). \label{eq:APQ}
\end{equation}
We call these \emph{active generators} (see Sec.~\ref{sec:Taxonomy-Discussion}), but ``antisymmetric'' is equally appropriate, modulo that they are distinct from the Hamiltonian generators, which are also antisymmetric.  The $A$ generators span the $\bA$ subspace of $\bL$.

\begin{figure}[t!]
\includegraphics[width=0.5\columnwidth]{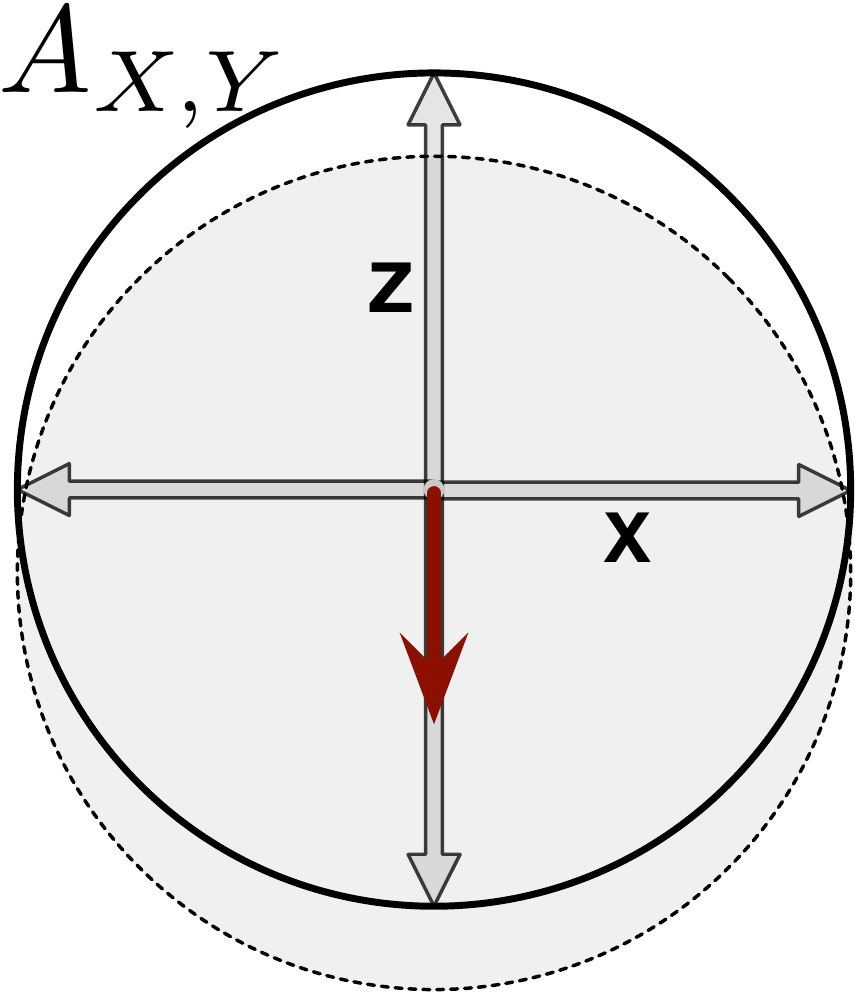}
\caption{The action of a representative active/antisymmetric error generator.  This is how the 1-qubit $A_{X,Y}$ generator acts on the $X\mdash Z$ plane of the Bloch sphere.  It has no action at all on the $X$, $Y$, or $Z$ operators, but sends $\Id \to -4Z$.  It thus shifts the entire Bloch sphere downward.}
\label{fig:AAction}
\end{figure}

\subsection{The dual basis}

We have constructed $d^2(d^2-1)$ linearly independent elementary error generators, in four classes, that partition the error generator space as $\bL = \bH \oplus \bS \oplus \bC \oplus \bA$.  Their collected actions, for easy reference, are:
\begin{align*}
H_P[\rho] &= -i[P,\rho] = -iP\rho \Id + i\Id\rho P,\tag{\ref{eq:HP}} \\
S_P[\rho] &= P\rho P - \Id\rho\Id,\tag{\ref{eq:SP}}\\
C_{P,Q}[\rho] &= P\rho Q + Q\rho P - \frac12\left\{\left\{P,Q\right\},\rho\right\},\tag{\ref{eq:CPQ}}\\
A_{P,Q}[\rho] &= i\left( P \rho Q -  Q\rho P + \frac{1}{2}\left\{\left[P,Q\right],\rho\right\}\right). \tag{\ref{eq:APQ}}
\end{align*}
These elementary error generators are not mutually orthogonal in the Hilbert-Schmidt inner product, but they have a very simple dual basis that can be used to extract each elementary generator's coefficient -- or \emph{rate} -- from an arbitrary error generator vector $\cL$:
\begin{eqnarray}
H'_{P}[\cdot] &=& -\frac{i}{d^2}[P,\cdot] = \frac{1}{d^2}H_P\\
S'_{P}[\cdot] &=& \frac{1}{d^2} P \cdot P \\
C'_{P,Q}[\cdot] &=& \frac{1}{2d^2}(P \cdot Q + Q \cdot P) \\
A'_{P,Q}[\cdot] &=& \frac{i}{d^2}\left(P \cdot Q - Q \cdot P\right).
\end{eqnarray}
These ``dual elementary generators'' are mutually orthogonal, and simpler than the elementary generators constructed above.  It's reasonable to ask why we didn't just start with them!  The answer is that they aren't trace-preserving (TP).  Making the elementary generators trace-preserving requires mixing in the special symmetric Choi units $X_{P,\Id} + X_{\Id,P}$, which never appear in the dual elementary generators.

We begin to explore the properties and nature of the $H$-, $S$-, $C$-, and $A$-type error generators by constructing them explicitly for 1- and 2-qubit systems as a concrete example.

\subsection{1-qubit elementary error generators} \label{sec:OneQubit}

\begin{figure*}[t!]
\includegraphics[width=2\columnwidth]{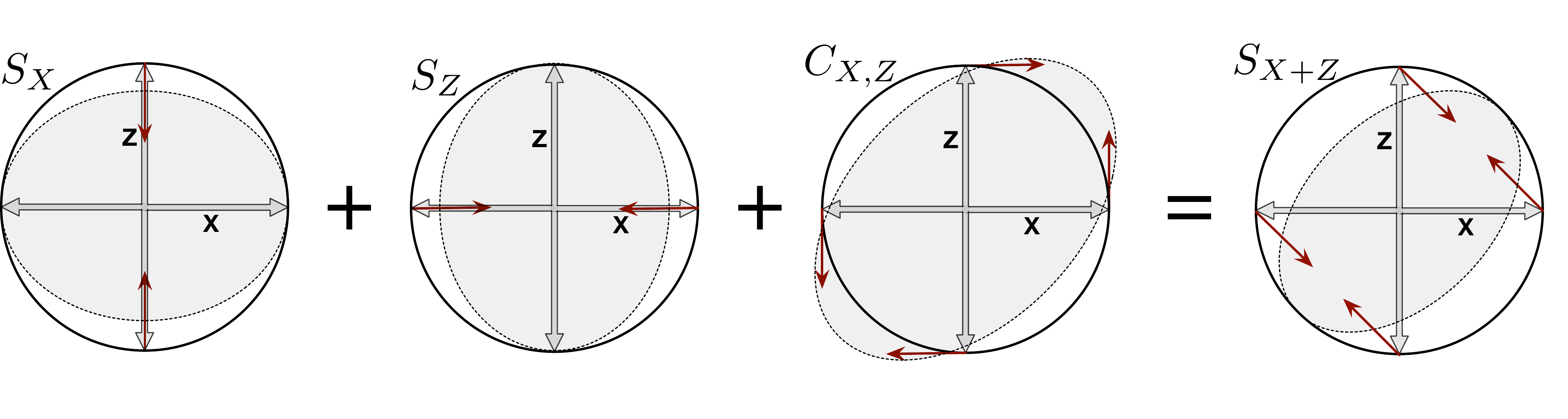}\label{fig:SXZ-decompose}
\caption{Pauli correlation generators are ``modifiers'' that combine with stochastic generators to generate non-Pauli stochastic errors.  As shown here, the sum $S_X + S_Z + C_{X,Z}$ generates a CPTP stochastic error process (dephasing) along the $X+Z$ axis.}
\end{figure*}

An arbitrary 1-qubit CPTP process $G$ that is close to its target $\bar{G}$ can be described by a $4\times 4$ superoperator, written in the Pauli basis $(\Id,X,Y,Z)$, whose top row is fixed to ensure trace preservation \cite{Ruskai2002-lk}:
\begin{equation}
    G = \left(\begin{array}{cccc}1&0&0&0\\a&b&c&d\\e&f&g&h\\j&k&l&m\end{array}\right)\bar{G}.
\end{equation}
This error process has 12 free parameters, which map to 12 elementary error generators:
\begin{itemize}
\item 3 Hamiltonian generators indexed by a Pauli $P$ ($H_X, H_Y, H_Z$),
\item 3 Pauli-stochastic generators indexed by a Pauli $P$ ($S_X, S_Y, S_Z$),
\item 3 Pauli-correlation generators indexed by nonequal pairs of Paulis $P,Q$ ($C_{X,Y}, C_{Y,Z}, C_{X,Z}$), and 
\item 3 active generators indexed by nonequal pairs of Paulis $P,Q$ ($A_{X,Y}, A_{Y,Z}, A_{X,Z}$).
\end{itemize}
Any single-qubit error generator $\cL$ can be written as a linear combination of these generators (Eqs. \ref{eq:LHSCA}-\ref{eq:Lsum}).

$H_P$ terms generate unitary rotations of the Bloch sphere, which are coherent errors in the gate $G$ (see Fig.~\ref{fig:HAction}).  The three coefficients $(h_X,h_Y,h_Z)$ indicate the rate of erroneous rotation with respect to each Pauli axis.  Together they form a Bloch sphere vector $\vec{h}$ whose direction is the axis of the unitary rotation, and whose length is its angle\footnote{Technically, $\vec{h}$ is a \emph{pseudovector}, since it defines a rotation.  The affine rates, discussed a bit lower down, form a true vector.  This distinction is irrelevant as long as everyone involved has the courtesy and common sense to avoid performing antiunitary coordinate changes.}.  If $G$'s error is purely coherent, then its error generator will be restricted to the $\bH$ subspace.

$S_P$ terms shrink the Bloch sphere to an ellipsoid aligned with the $X/Y/Z$ axes (see Fig.~\ref{fig:SAction}).  Each $S_P$ generates dephasing toward the $P$ axis -- e.g., $S_Z$ shrinks polarization along the $X$ and $Y$ axes, but leaves $\expect{Z}$ alone.  The three $S_P$ generators generate Pauli channels, which appear in quantum error correction theory \cite{Eastin2007-kq, Flammia2011-rw, Chiuri2011-os, Gutierrez2013-ey, Gutierrez2015-qe, Tuckett2018-bb, Beale2018-ms, Huang2019-ua} and quantum process characterization \cite{Flammia2019-gv, Harper2020-np}.  

Dephasing along other axes in the Bloch sphere is also possible.  It is generated by \emph{combinations} of the $S_P$ and the Pauli correlation $C_{P,Q}$ generators (Fig.~\ref{fig:SXZ-decompose}).  By themselves, the three $C_{P,Q}$ generators are never physically valid (they generate non-CP maps).  They generate zero-determinant ``squeezing'' of the Bloch sphere, causing it to shrink and grow along perpendicular axes.  For example, the $C_{X,Z}$ generator acts (see Fig.~\ref{fig:CAction}) as
\begin{eqnarray*}
C_{X,Z}[X+Z] &=& X+Z, \\
C_{X,Z}[X-Z] &=& -(X-Z), \\
C_{X,Z}[Y] &=& C_{X,Z}[\Id] = 0,
\end{eqnarray*}
so it ``inflates'' the $X+Z$ axis of the Bloch sphere and shrinks the $X-Z$ axis.  But adding it to $S_X + S_Z$ yields a generator of $X+Z$ errors that preserves $X+Z$ and shrinks $X-Z$, dephasing the Bloch sphere toward the $X+Y$ axis (see Fig.~\ref{fig:SXZ-decompose}).  

Unlike the Hamiltonian rates $(h_X,h_Y,y_Z)$, which form a Bloch sphere \emph{vector}, the stochastic rates form a symmetric \emph{tensor} in the same space,
\begin{equation}
    \Sigma = \left(\begin{array}{ccc} s_X & c_{X,Y} & c_{X,Z} \\ c_{X,Y} & s_Y & c_{Y,Z} \\ c_{X,Z} & c_{Y,Z} & s_Z \end{array}\right),
\end{equation}
which describes an error process that shrinks the Bloch sphere to an ellipsoid.

\begin{figure*}[t!]
\includegraphics[width=2\columnwidth]{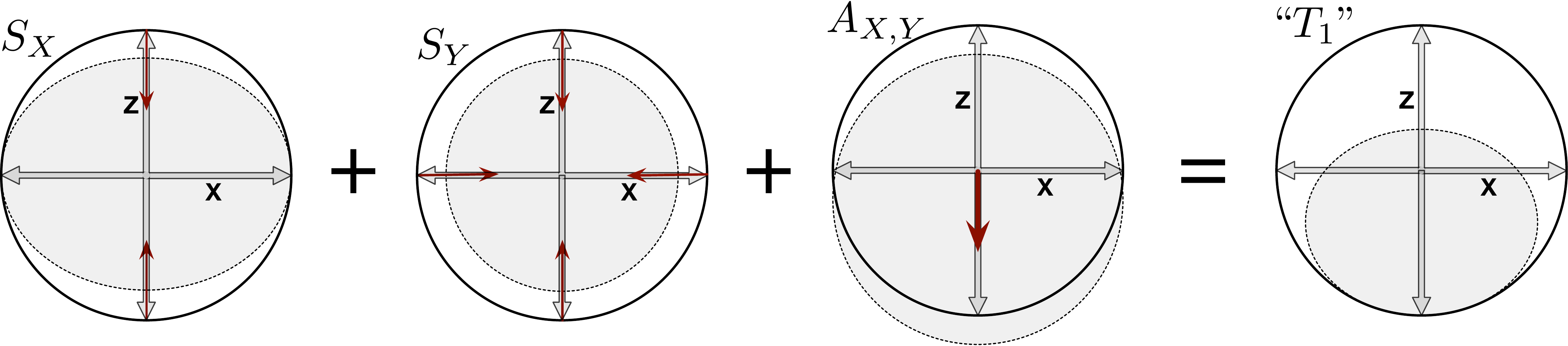}
\caption{Like Pauli-correlation ($C$) generators, active/antisymmetric generators are ``modifiers'' that combine with stochastic generators to generate well-known processes.  As shown here, the sum $S_X + S_Y + A_{X,Y}$ generates a non-unital amplitude-damping or ``$T_1$'' process that describes cooling or dissipative decay toward the $\proj{0}$ state.}
\label{fig:T1-decompose}
\end{figure*}

The three $A_{P,Q}$ generators, which we called ``active'' or ``antisymmetric'', can also be called ``affine'' generators for a single qubit, because they generate affine shifts of the Bloch sphere.  For example, $A_{X,Y}$ acts (see Fig.~\ref{fig:AAction}) as
\begin{eqnarray*}
A_{X,Y}[\Id] &=& -4Z \\
A_{X,Y}[X] &=& A_{X,Y}[Y] = A_{X,Y}[Z] = 0.
\end{eqnarray*}
So for \emph{any} density matrix $\rho$, $A_{X,Y}[\rho] = -2Z$.  Similarly, $A_{X,Z}$ and $A_{Y,Z}$ generate affine shifts in the $Y$ and $X$ directions, respectively.  Like the $C$-type generators, these are never physically valid by themselves.  But when combined with $S$-type generators, they produce non-unital decoherence processes.   The best-known example is $T_1$ decay, a.k.a. amplitude damping from $\ket{1}$ to $\ket{0}$, given by
\begin{equation} \label{eq:AmpDamp}
\Gamma_{1\to0}[\rho] = \gamma \sigma_-\rho\sigma_-^{\dagger} + \sigma_0\rho\sigma_0^\dagger
\end{equation}
where $\sigma_- = \ketbra{0}{1}$ and $\sigma_0 = (\proj{0}+\sqrt{1-\gamma}\proj{1})$.  Amplitude damping shrinks the Bloch sphere to a Pauli-aligned ellipsoid (like a combination of Pauli-stochastic errors), but also shifts it affinely in the $Z$ direction.  Its generator (see Fig.~\ref{fig:T1-decompose}) is proportional to
\begin{equation}
S_X + S_Y + A_{X,Y}.    
\end{equation}
The process given in Eq.~\ref{eq:AmpDamp} is often treated as a wholly independent kind of error, distinct from stochastic Pauli errors.  It's reasonable to ask why we represent it as a combination of stochastic Pauli errors and a non-CP affine shift, rather than as an independent elementary error generator.  The answer is simple:  there are \emph{too many} processes like the one given in Eq.~\ref{eq:AmpDamp}!  

Given any point on the Bloch sphere, a process can be constructed that ``damps'' toward it\footnote{It can be constructed as $\mathcal{U} \Gamma_{1\to0} \mathcal{U}^\dagger$ where $\mathcal{U}[\cdot] = U\cdot U^{\dagger}$ for a suitable unitary $U$.}.  But our goal here is to construct a set of linearly independent generators, so that any $L$ can be uniquely decomposed as a linear combination of them.  Those ``damping'' processes cannot all be linearly independent.  But if we consider them in the context of the $S$ and $C$ generators already constructed, then the $A$ generators precisely span the additional degrees of freedom contributed by ``damping'' processes.
For example, the $A_{X,Y}$ generator is a \emph{difference} of amplitude-damping processes,
\begin{equation}
A_{X,Y} \propto \Gamma_{1\to0} - \Gamma_{0\to1}.
\end{equation}
The $A_{X,Z}$ and $A_{Y,Z}$ affine generators can be defined similarly in terms of amplitude-damping processes in the $Y$ and $X$ bases, respectively.  Their coefficients $(a_{y,z}, a_{x,z}, a_{x,y})$ form a Bloch sphere vector that indicate the direction in which the maximally mixed state will be shifted by the $\bA$ portion of the error generator.

\subsection{Elementary error generators for 2 qubits} \label{sec:TwoQubit}

Many features and interpretations from the 1-qubit example carry over to a 2-qubit system.  However, there are a few novelties that make it worth examining.

If we consider a 2-qubit process $G$ and its error generator $\cL$ (Fig.~\ref{fig:2Q-Partition}), they both have $256-16=240$ free parameters.  We can identify and construct Hamiltonian and Pauli-stochastic generators (15 each, indexed by the 15 nontrivial 2-qubit Pauli operators) exactly as for 1 qubit.  

There are 105 linearly independent 2-qubit Pauli correlation generators.  A new phenomenon appears for 2 qubits, because $P\rho Q + Q\rho P$ is only TP if $P$ and $Q$ anticommute.  In the 1-qubit example, every pair of distinct traceless Pauli operators anticommutes.  But in a 2-qubit system, there are also pairs such as $(IZ,ZZ)$ that commute.  For these cases, the double anticommutator term in the definition of $C_{P,Q}$ (Eq.~\ref{eq:CPQ}) becomes nonzero and ensures that the generator is trace-preserving.  This term has some odd consequences, discussed in Section \ref{sec:Metrics}.

The $A$-type generators hold more surprises.  In the 1-qubit example, there were exactly 3 $A$-type generators.  They all generated pure affine shifts (e.g. $A_{X,Y}[\rho] = -2Z$), and their rates conveniently formed a vector in Hilbert-Schmidt space.  But for a 2-qubit system, there are 105 $A$-type generators, and they do not generally produce affine shifts in any particular direction.

To understand their action, we start by considering $A_{P,Q}$ when $P$ and $Q$ both act nontrivially only on one (the same) qubit -- e.g., $A_{X\Id,Y\Id}$.  This elementary generator can also be written as $A_{X,Y}\otimes \Id$.  It generates an affine shift on qubit 1 \emph{independent} of qubit 2's state:
\begin{eqnarray}
    A_{X\Id,Y\Id}[\rho_1\otimes\rho_2] &=& -2Z\otimes\rho_2,\\
    A_{X\Id,Y\Id}[\rho_{12}] &=& -2Z\otimes\Tr_1(\rho_{12}).
\end{eqnarray}
It naturally appears as part of the generator for a local amplitude damping process $\Gamma_{1\to0}\otimes\Id$ that acts only on qubit 1.

This example can be generalized.  It involved two anticommuting Paulis acting on one qubit.  But any two anticommuting Paulis generate a 1-qubit Pauli algebra, and can be viewed as the effective ``$X$'' and ``$Y$'' operators for a virtual qubit encoded somewhere within the 2-qubit Hilbert space.  So, every $A_{P,Q}$ where $\{P,Q\}=0$ induces an affine shift in the direction of $i[P,Q]$ on \emph{some} virtual 1-qubit subsystem.  For example, both $A_{X\Id,Y\Id}$ and $A_{XX,YX}$ induce affine shifts in the $Z\Id$ direction, but they do so acting on different 1-qubit subsystems.

There are also $A$-type generators that produce no affine shift at all.  One such example is a difference of two $A$-type generators that produce the same affine shift, eg. $A_{X\Id,Y\Id} - A_{XX,YX}$.  A more fundamental example, though, is given by the $A_{P,Q}$ generators for \emph{commuting} pairs of Paulis. Here, $[P,Q]=0$, and $A_{P,Q}[\rho] = i(P\rho Q - Q\rho P)$, so $A_{P,Q}[\Id] = 0$.  No affine shift occurs -- the generated process is unital.

We do not have a clear or intuitive understanding of these generators, and when (if ever) they are likely to have significant rates in realistic quantum processors.  However, we discuss their action, and the rationale for calling them ``active'', in the next section.

\subsection{Discussion} \label{sec:Taxonomy-Discussion}

A taxonomy is a classification, of things or concepts.  Our main accomplishment to this point is showing \emph{how} to represent the small Markovian error in a gate $G$ as an error generator $L$, and to decompose $L$ in a basis of elementary error generators.  We now return to the implicit promise of the title, and present a taxonomy of small Markovian errors that classifies them into sensible categories.  Most of this analysis is implicit in the preceding sections, so our goal here is to make it explicit and fill in the gaps.

We cannot classify an arbitrary $L$ as representing one kind of error \emph{or} another, because it will be a linear combination of all the elementary generators listed above.  Instead, $L$ represents a \emph{mixture} of different sorts of errors.  The elementary generators constitute a classification of the phenomena that can be mixed together to make an arbitrary error $L$.  So in analyzing such an $L$, the taxonomy enables separating those components out to make statements like ``The error in $g$ is primarily Hamiltonian (coherent),'' or ``The error in $g$ is 35\% Hamiltonian and 65\% stochastic, and the stochastic error is almost all Pauli-stochastic,'' or ``The rate of single-qubit Hamiltonian errors in $g$ is 1.7\%.''  We seek to complete the project that Kueng \emph{et al} \cite{Kueng2016-vy} initiated by classifying single-qubit error modes and their impact.

The root of the taxonomy we propose is the partition of error generator space ($\bL$) into Hamiltonian ($\bH$), stochastic ($\bS\oplus\bC$), and active ($\bA$) subspaces.  These subspaces are unitarily invariant, which means that the classification of a particular error as ``Hamiltonian'', ``stochastic'', or ``active'' is unaffected by time evolution, change of basis, or whether the generator comes before or after (or during) a unitary target gate.  So, given an arbitrary $\cL$, it's meaningful to separate it into its components on distinct sectors, $\cL = \cL_\bH + (\cL_{\bS}+\cL_{\bC}) + \cL_\bA$, and consider each term (more or less) independently.  As belied by our notation, we often find it useful to subdivide the stochastic sector into Pauli-stochastic and Pauli-correlation.  This division is less fundamental, as we discuss below.

\textbf{Hamiltonian errors} are widely well-understood and require little discussion.  Any $\cL_\bH$ whatsoever generates a legitimate and physically valid unitary error process, often called a \emph{coherent error}.  Coherent errors can be eliminated, in principle, by dynamical decoupling or recalibrating control.  Variations of $\cL_\bH$ are completely independent of errors on the other sectors -- e.g., changing $\cL_\bH$ does not affect the complete positivity or the interpretation of any other error generators.

\textbf{Stochastic errors} are unital (they preserve the maximally mixed state), and produce no net rotation of the state space.  We introduced them as the non-unitary consequences of small \emph{random unitary} dynamics -- i.e., mixtures of different unitary rotations \cite{Audenaert2008-ra}.  But like most quantum processes, stochastic errors can have multiple distinct physical causes\footnote{So inferring the precise cause of a given error generator is generally not justified.}.  Stochastic errors can also be produced by \emph{minimally disturbing measurements} \cite{Busch2009-lu} whose outcomes are ignored.  For example, if a qubit's environment measures it weakly in the $Z$ basis, the qubit will experience dephasing, which is a stochastic $Z$ error.  More generally, discarding the outcome of any minimally disturbing measurement produces a quantum process
\begin{equation}
    \cE[\rho] = \sum_k{ M_k \rho M_k}, \label{eq:MinimalMeasurement}
\end{equation}
where each $M_k\geq 0$ and $\sum_k{M_k^2}=\Id$.  Positivity implies that each $M_k$ is Hermitian, and can be expanded as a sum of Paulis with strictly real coefficients.  This means the error process in Eq. \ref{eq:MinimalMeasurement} has a strictly symmetric Choi sum form, so it is orthogonal to all $H$- and $A$-type generators, and thus wholly stochastic.

There is no compelling \emph{mathematical} reason to separate the $\bS$ and $\bC$ generators.  As observed earlier, their rates are the diagonal and off-diagonal components of a symmetric tensor, and so unitary changes of basis will mix them.  But separating them makes \emph{practical} sense, because the Pauli basis is very special in quantum information science.  Experimental gates are often generated by Pauli Hamiltonians.  Stochastic Pauli error rates appear throughout the theory of quantum error correction \cite{Eastin2007-kq, Flammia2011-rw, Chiuri2011-os, Gutierrez2013-ey, Gutierrez2015-qe, Tuckett2018-bb, Beale2018-ms, Huang2019-ua, Flammia2019-gv, Harper2020-np}.  And the Clifford group -- the automorphism group of the Paulis -- appears frequently in both experimental gate sets and theoretical constructions \cite{Aaronson2004-im, Gutierrez2013-ey, Zhu2016-xm, Magesan2011-ra, Dankert2009-uu}.  Although unitary transformations do not generally preserve the $\bS$/$\bC$ separation, Clifford transformations do.  So, in an experimental or theoretical context that privileges the Pauli or Clifford operations, this separation is likely to be useful.  In others, it may not.

There are two key distinctions between the $S$ and $C$ generators.  First, any linear combination of $S$ generators yields a physically valid process if (and almost\footnote{See Sec.~\ref{sec:Lindblad} for a more careful and context-aware discussion of this constraint.} only if) the $s_P$ rates are all non-negative.  In contrast, the rate of each $C_{P,Q}$ can be either positive or negative, but it is strictly bounded by the rates of $S_P$ and $S_Q$.  The constraint is nontrivial but easy to state: the symmetric matrix with diagonal elements $S_P$ and off-diagonal elements $C_{P,Q}$ must be positive semidefinite.  A simple consequence is that $|c_{P,Q}| \leq \sqrt{s_Ps_Q}$.  This has useful sparsity implications for estimating error processes: if only $n$ of the $d^2-1$ Pauli-stochastic error rates are non-negligible, then all but $n(n-1)/2$ Pauli correlation rates can also be neglected.  

Second, while $S$ generators \emph{cause} errors -- i.e., their rates really are ``error rates'' of bit- or phase-flip errors -- $C$ generators only \emph{modify} them.  For example, if a qubit has $s_X = s_Z = 0.01$, then the total rate of bit- and phase-flip errors is 0.02.  Varying the $c_{X,Z}$ coefficient over its entire range doesn't change this.  It merely shifts the error mechanism from pure dephasing in the $X-Z$ basis, to independent bit- and phase-flip errors, to pure dephasing in the $X+Z$ basis.  

\begin{figure}[t!]
\includegraphics[width=1\columnwidth]{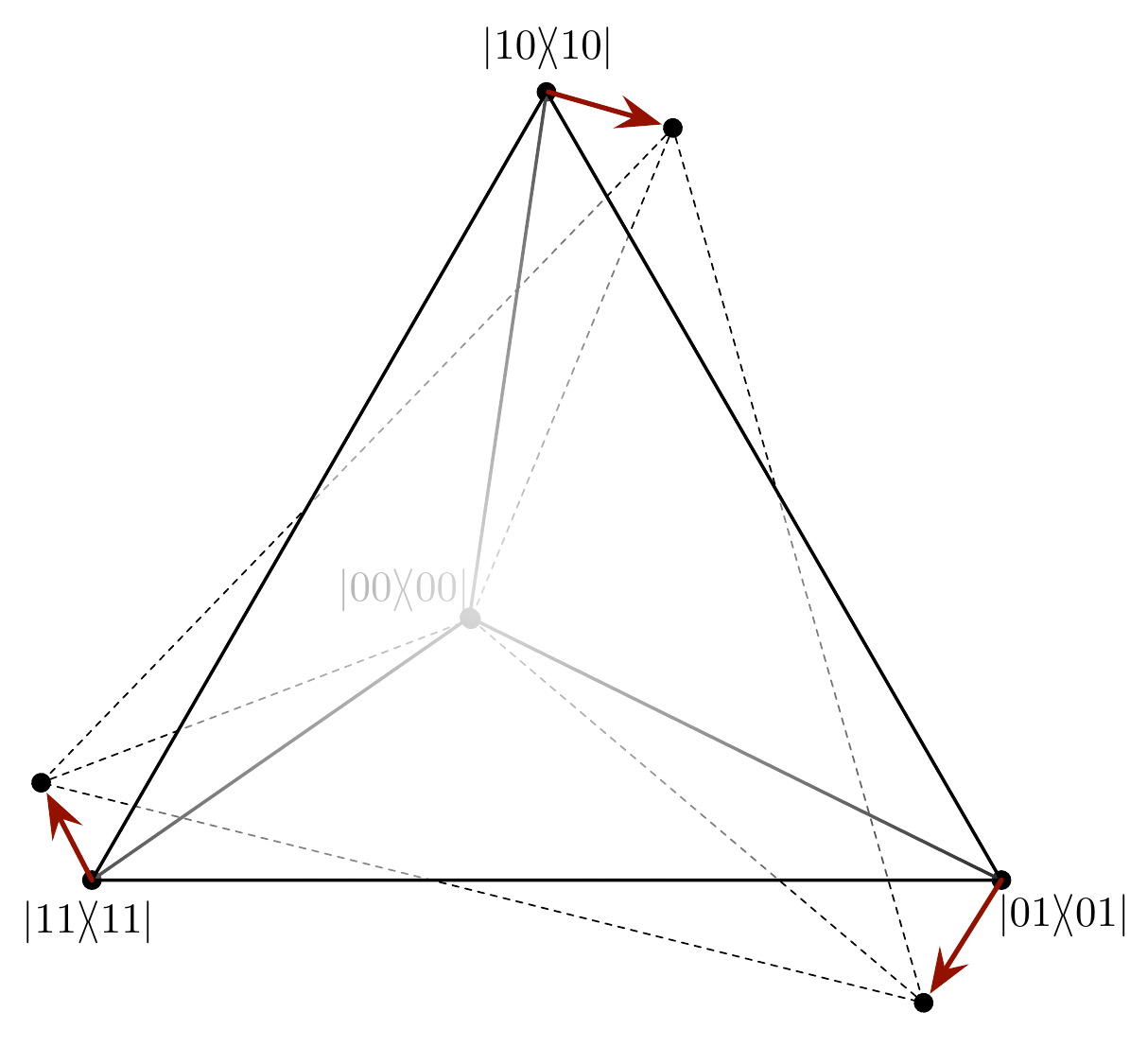}
\caption{$A$-type generators, by themselves, are always non-CP.  Some of them generate affine shifts on 1-qubit virtual subsystems, but in general $A$-type generators cause orthogonal $SO(d^2-1)$ rotations on the space of density matrices that do \emph{not} correspond to unitary $U(d)$ transformations on Hilbert space.  A 2-qubit example, shown here, is a rotation entirely within the space spanned by the projectors onto the 4 computational basis states -- or, equivalently, by $\{\Id\Id, \Id Z, Z\Id, ZZ\}$.  The convex hull of the 4 computational basis states defines a classical simplex that is a cross-section of the quantum state space.  This $SO(4) \subset SO(15)$ rotation is generated by a linear combination of $A$-type generators, and it maps the vertices of the simplex to points outside the simplex, which are not positive semidefinite density matrices.}
\label{fig:NonCP-active}
\end{figure}

\textbf{Active ($A$-type) errors} are relatively mysterious.  We do not understand them to the same degree as Hamiltonian or stochastic errors.  The effects that they produce appear rarely in theoretical models of quantum errors -- with the sole major exception of $T_1$ (amplitude-damping) processes.  Amplitude damping is non-unital (it can decrease entropy), and only $A$-type error generators produce non-unital error processes.

But $T_1$ decay, in quantum computing, is usually a single-qubit effect.  It's a cooling process, where weak coupling between the system's dominant Hamiltonian and a large cold environmental bath produces irreversible decay into lower-energy states.  Gate-model quantum processors are usually designed with steady-state Hamiltonians that do not couple the qubits, and transient controllable coupling Hamiltonians that aren't ``on'' long enough for qubits to cool into correlated ground states.  Independent $T_1$ decay on isolated qubits can be modeled entirely just the weight-1 $A_{P,Q}$ generators where $P$ and $Q$ act trivially on all but one qubit.  Which leaves open the question of what all the other $A$ generators do.

Part of the answer is that multiqubit systems can have complicated Hamiltonians, and experience complicated cooling processes.  For example, modeling ``$T_1$'' decay for a 2-qubit system with a single ground state and a triply degenerate excited state (call it $\Gamma_{\{1,2,3\}\to0}$) would require additional $A$-type generators.  But this can't be the whole story, because the $A_{P,Q}$ generators for \emph{commuting} $(P,Q)$ aren't non-unital at all, so they have no direct relationship to cooling.  These generators are entirely antisymmetric. So, like Hamiltonian generators, they generate orthogonal $SO(d^2-1)$ rotations that rotate $\rho$ in $\cL(\cH)$ without changing $\Tr\rho^2$.  But these rotations that do \emph{not} correspond to unitaries, and are therefore not CP by themselves (see Fig.~\ref{fig:NonCP-active}).

The most general statement we can make about the $A$-type generators is that they are all produced by some form of \emph{active feedback} from the environment\footnote{This (finally) is the motivation for the ``active'' moniker.}.  Any quantum process can be written using a diagonalized Choi sum form known as the \emph{Kraus representation} \cite{Leung2003-az},
\begin{equation}
    \cE[\rho] = \sum_k{K_k \rho K_k^\dagger},
\end{equation}
subject only to the constraint $\sum_k{K_k^\dagger K_k = \Id}$.  If we use the polar decomposition to write each $K_k = U_k M_k$, where $U_k$ is unitary and $M_k$ is positive semidefinite, then it's clear that $\cE$ can be implemented by (1) performing a minimally disturbing measurement of the POVM  $\{M_k^2\}$, (2) applying $U_k$ \emph{conditional} on the observed result, and (3) discarding the observed outcome $k$.  If $U_k = \Id$ for all $k$, then the process results from a minimally disturbing measurement.  We showed earlier that such processes are entirely modeled by stochastic error generators.  If $U_k = \mathrm{constant}$ regardless of $k$, then the process is the composition of (1) a minimally disturbing measurement and (2) a unitary.  This corresponds to a combination of Hamiltonian and stochastic errors.  It follows that $A$-type error generators are uniquely associated with error processes where $U_k$ depends on $k$.  This constitutes active feedback; the environment induces system dynamics \emph{conditional} on the result of a measurement on the system.

Cooling (a.k.a. $T_1$ or amplitude-damping) is a special and well-understood example of active feedback.  A $T_1$ process can be written explicitly as a one-sided weak measurement of $Z$, $\{p\proj{1}, \Id-p\proj{1}\}$, followed by a bit-flip ($X$) operation conditional on observing the excited state $\ket{1}$.  But active feedback can produce many other dynamics too.  If we make $\{M_k^2\}$ a weak version of an informationally complete POVM, then by choosing the conditional unitaries $U_k$, we can produce a component of literally \emph{any} orthogonal rotation in $SO(d^2-1)$.

\begin{figure}[t!]
\includegraphics[width=1\columnwidth]{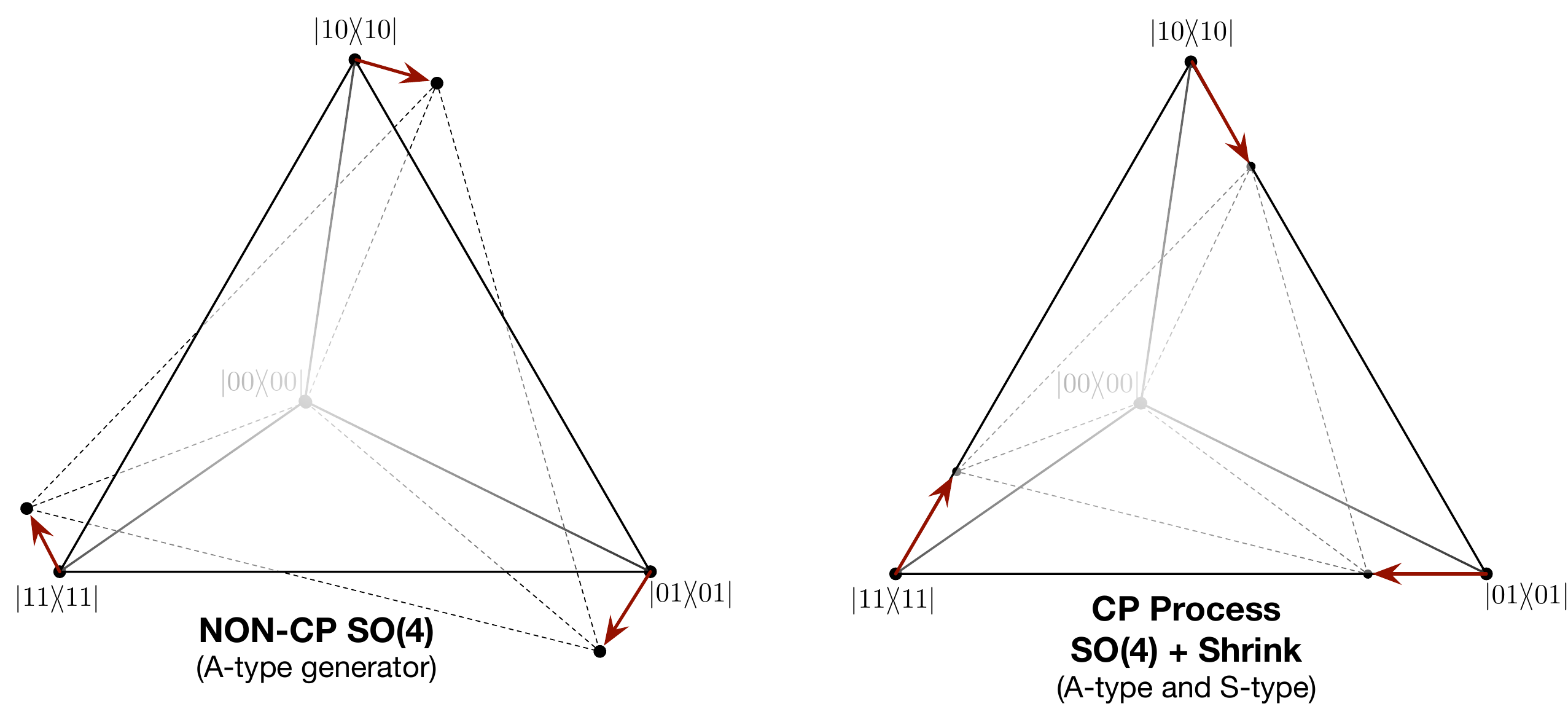}
\caption{When maps generated by $A$-type generators are implemented via non-minimally-disturbing measurements (measurement followed by a conditional unitary), they are accompanied by $S$-type decoherence that shrinks the state space enough to ensure complete positivity.  Here, this is shown for the example of Fig \ref{fig:NonCP-active}.  The illustration on the left shows the action of a non-CP process generated by \emph{only} $A$-type generators on the computational basis simplex.  The illustration on the right shows the action of a CPTP process with the same $A$-type generators, but additional $S$-type generators that are inevitable consequences of the measurement.  The $T_1$ process in Fig. \ref{fig:T1-decompose} can be seen as an example of exactly the same phenomenon, but with an $A$-type generator that produces an affine shift.}
\label{fig:CP-active}
\end{figure}

In this construction, the orthogonal rotation generated by the $A$ generators will be accompanied by a significant amount of stochastic error ($S$ generators), which is caused by the measurement, and not eliminated by the conditional unitary (see Fig.~\ref{fig:CP-active}, and also Fig.~\ref{fig:T1-decompose} in retrospect).  This illustrates that, like the Pauli correlation generators, $A$-type generators are ``modifiers''.  They are never completely positive by themselves.  The rate of $A_{P,Q}$ can have either sign, but cannot be nonzero unless the corresponding $S_P$ and $S_Q$ error rates are both nonzero, and $|a_{P,Q}| \leq \sqrt{s_Ps_Q}$.  Their main role is not to \emph{create} errors, but to move their impact around.  The single-qubit $T_1$ process provides a good example of this.  Recall that it is generated by $S_X + S_Y + A_{X,Y}$.  The stochastic error process generated by $S_X+S_Y$ alone flips both the $\ket{0}$ and $\ket{1}$ states with equal probability $p$.  The $T_1$ process, on the other hand, never flips the $\ket{0}$ state, but flips the $\ket{1}$ state with probability $2p$.  This difference can be significant -- for example, transmon-based processors perform better when ancilla qubits are stored in the $\ket{0}$ state, because of precisely this effect -- but it does not change the \emph{average} error rate of the gate.

\section{Relationship to Lindblad master equations}\label{sec:Lindblad}

We defined an error generator $\cL$ as the logarithm of an error process $\cE$ that is close to the identity process.  This is very similar to -- but not quite the same as -- the generator of a Lindblad master equation \cite{Lindblad1976-qr}, or \emph{Lindbladian}.

A superoperator $\Lind$ is a Lindbladian iff $e^{t\Lind}$ is a CPTP map for all $t>0$.  So for every Lindbladian $\Lind$ there is a corresponding error process $\cE = e^{\Lind}$ whose error generator $L = \log\cE$ is equal to $\Lind$.  But the reverse does not hold:  there are error processes $\cE$ whose error generator $L = \log\cE$ is not a Lindbladian.  These processes are called \emph{non-infinitely-divisible} processes\footnote{This is just one stage in a beautiful hierarchy of divisibility conditions outlined in Ref. \cite{Wolf2008-hx}.  There are also \emph{indivisible} processes, which cannot be nontrivially decomposed at all into $\cE = \cE_1\circ\cE_2$.}, and are sometimes called \emph{non-Markovian} (see, e.g., Ref. \cite{Wolf2008-mi}).  This nomenclature can produce confusion here, since we say that \emph{any} process describes Markovian dynamics.  

The root cause of the confusion is the distinction between continuous-time dynamics, and discrete-time dynamics.  A process has the Markov property if (and only if) its state at time $t'>t$ is determined entirely by the state at time $t$ (and the nature of the process, of course).  It is necessary to state what values $t$ may take.  In contexts where $t$ takes ordinal values (e.g. $t=0,1,2,\ldots$), time is discrete; when $t$ is allowed to take real values, time is continuous.

So, \emph{any} quantum process (CPTP map) $\cE$ can define a discrete-time Markov semigroup,
\begin{equation}
    \{\Id, \cE, \cE^2, \cE^3,\ldots\}.
\end{equation}
But it is reasonable to go further and ask whether this discrete semigroup can be embedded in a continuous-time semigroup $\{e^{t\log\cE}\}$.  This is equivalent to asking whether $\cE$ could have been \emph{caused} by continuous-time Markovian dynamics.  In contrast, the discrete-time semigroup describes the discrete-time Markovian dynamics that \emph{results} from $\cE$.  Both are interesting and useful concepts.  But the continuous-time paradigm is more relevant to understanding the causes and mechanisms of error within a gate, while the discrete-time paradigm is more relevant to understanding the computational consequences of those errors.  (A pithy summary might be ``Physics clocks are continuous; computer clocks are discrete.'')

So, in this article, we do not intrinsically care about the divisibility of $\cE$.  In fact, we explicitly avoided any inference about how $G$ was generated (see Sec. \ref{sec:Generators}), because quantum logic gates are \emph{always} induced by time-varying processes, and so we do not generally expect them to be consistent with Markovian continuous-time dynamics.  But this doesn't provide license to ignore divisibility, because if $\cE$ is \emph{not} infinitely divisible, then its logarithm $\cL$ is not a Lindbladian.  This has consequences.  Our goal in this section is to discuss, bound, and mitigate those consequences.

If $\cL = \log\cE$ is not Lindbladian, then there exists \emph{some} $t$ for which $e^{t\cL}$ is not CP.  Since $\cE = e^{\cL}$ is CP, violations of CP can never be observed at integer $t$, and the largest violation will occur for $t\in(0,1)$.  We do not propose to construct such maps; our goal is to describe errors in $G$, and to compute their consequences when a discrete (integer) number of gates are applied in sequence.  Within this context, we will never observe violations of CP directly.

What we \emph{do} need to deal with, however, are the analytic consequences of $\cL$ not (necessarily) being a Lindbladian.  These consequences are mitigated by our restriction to \emph{small} errors -- i.e., to $\cE \approx \Id$.  Indivisible maps can be found arbitrarily close to $\Id$, so small errors don't eliminate the issue.  A useful example of an indivisible map close to $\Id$ is the process whose action is
\begin{equation}
\cE[\rho] = (1-2p)\rho + p X\rho X + p Y\rho Y. \label{eq:example-indivisible}
\end{equation}
A Lindblad generator for this process would need to independently generate $X$ and $Y$ errors at rate $p$.  But such a generator would produce an $X$ error \emph{and} a $Y$ error with probability $p^2$, and this effects a $Z$ error.  No continuous-time Markovian process can cause both $X$ and $Y$ errors without also causing $Z$ errors.  If we compute $\log\cE$ for Eq.~\ref{eq:example-indivisible}, we get an $O(p^2)$ \emph{negative} rate of $Z$ errors, which makes the net probability of a $Z$ error equal to zero at $t=1$, but makes $\exp(t\log\cE)$ non-CP for $t \in (0,1)$.

Restricting to small errors mitigates this issue because, although $\cL$ is not always a Lindbladian, it is very \emph{close} to a Lindbladian when the error is small.  More precisely:  if $\cL = \log\cE$, then there exists a valid Lindbladian $\cL'$ such that $\cL' - \cL = O(\cL^2)$.  To show this, it is sufficient to choose $\cL' = \cE - \Id$.  This is a valid Lindbladian (Lemma 1 of \cite{Wolf2008-hx}), and its closeness to $\bL$ follows from the series expansion $\log(\cE) = (\cE-\Id) - \frac12(\cE-\Id)^2 + o\left((\cE-\Id)^2\right)$.  A useful corollary is that although $\cE \approx \Id$ may not be divisible, there always exists a nearby divisible $\cE'$ with $|\cE'-\cE| = O(|\cE-\Id|^2)$.

These two observations provide two ways to use $\cL$ as ``almost'' a Lindbladian.  First, we can take the logarithm of any small $\cE$, and treat it as a Lindbladian, at the price of accepting small $O(\cL^2)$ violations of positivity.  Second, we can treat $\cL$ as the fundamental thing, restrict it to be a valid Lindbladian, and be confident that we can approximate any $\cE$ to within $O(|\cE-\Id|^2)$ by $\cE\approx e^{\cL}$.

We could have avoided non-positivity entirely by simply defining $L$ differently, as $\cL = \cL^\Delta \equiv \cE - \Id$ in Sec.~\ref{sec:Generators}.  This is still a valid choice -- none of the subsequent analysis (e.g. elementary error generators) would be different, and various errors' rates would only change at $O(|\cL^2|)$.  A clean example of how this would change error rates can be obtained by considering two slightly different versions of the example given in Eq.~\ref{eq:example-indivisible}:
\begin{eqnarray}
\cE_{a} &=& \rho \to (1-2p)\rho + p X\rho X + p Y\rho Y,\\
\cE_{b} &=& \exp\left(pS_X + pS_Y\right).
\end{eqnarray}
Both are CPTP.  The first is indivisible; the second is continuous-time Markovian.  If we compute error generators using the logarithm, we get
\begin{eqnarray}
\cL_{a} &=& p S_X + p S_Y - p^2 S_Z + o(p^2),\\
\cL_{b} &=& p S_X + p S_Y.
\end{eqnarray}
If we compute them using the difference, we get
\begin{eqnarray}
\cL^\Delta_{a} &=& p S_X + p S_Y,\\
\cL^\Delta_{b} &=& p S_X + p S_Y + p^2 S_Z + o(p^2).
\end{eqnarray}
In our experience, we have found the first representation to be at least theoretically more informative -- nondivisible error processes are flagged by negative error rates, and the rates of rare error processes (e.g. correlated errors) are not contaminated by quotidian collisions between two common error processes.  But we believe that which choice is ``better'' remains, at the least, an open question.  It is possible, as experimental resolution into error processes increases, that both conventions will find specific practical applications.

\section{Simple error metrics} \label{sec:Metrics}

We have shown how to take the process matrix for an imperfect gate, transform it to a list (or vector) of error rates, partition those rates into distinct classes (subspaces), and interpret each one.  But that list of error rates can still be very long.  Often, it is desirable to condense a detailed description of the error process into one or two summary statistics.  Several error metrics exist for process matrices \cite{Gilchrist2005-pf}, of which the most commonly used (in recent years) are \emph{process fidelity} (with the target gate) and \emph{diamond norm distance} (to the target gate).

\begin{figure}[t!]
\includegraphics[width=1\columnwidth]{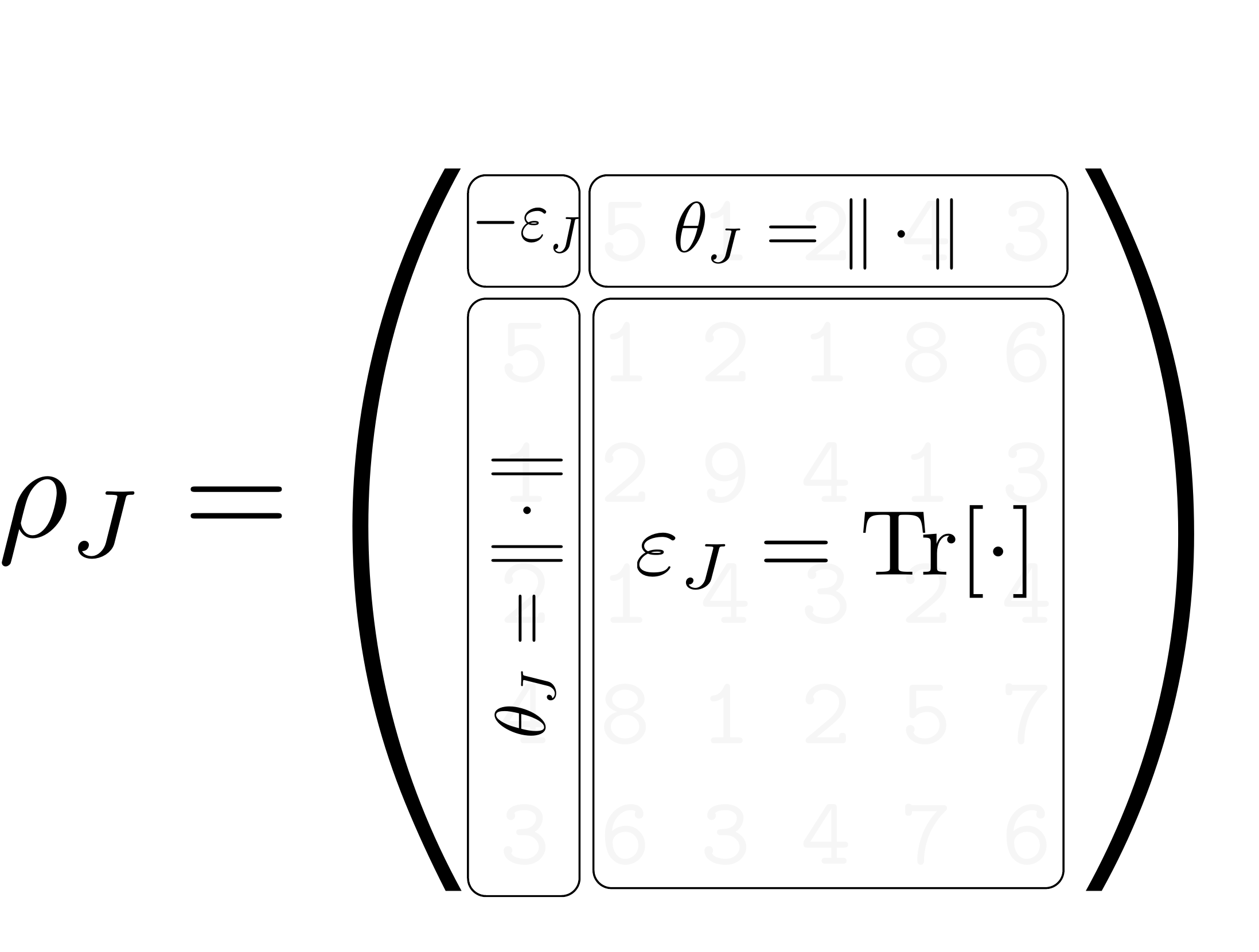}
\caption{The two simple error metrics that we introduce are both derived directly from the generator's \emph{Jamiolkowski state}, $\rho_J(\cL) = (\cL\otimes\Id)\left[\proj{\Psi}\right]$ (Eq. \ref{eq:Jamiolkowski}).  They quantify how a generator -- and the error process that it generates -- would impact a computation if they were applied to part of a maximally entangled state.  The Jamiolkowski probability ($\PJ$) measures the rate at which a generator transfers probability from the input state to its orthogonal complement.  It is the trace of the large diagonal block shown here.  The Jamiolkowski amplitude ($\theta_J$) measures how the rate at which the generator transfers amplitude from the input state to its orthogonal complement.  It is the norm of either of the off-diagonal blocks shown here.}
\label{fig:JamiolkowskiMetrics}
\end{figure}

Describing imperfect gates by their error generators does not preclude use of existing metrics. An error generator is a faithful description of the process matrix, so it's easy to reconstruct $G = e^{\cL}\bar{G}$ and compute its fidelity with, or diamond norm to, $\bar{G}$.  But error generators provide the opportunity to consider alternative metrics that are equally useful, and may be more natural for error generators.  Here, we introduce two such metrics, briefly discuss why they resonate with the error generator representation, and sketch their relationship to fidelity and diamond norm.  This is \emph{not} intended as a comprehensive investigation of the subject; that might be a good topic for future work.

Both metrics are based on the \emph{Jamiolkowski state} of a process \cite{Jamiolkowski1972-cf}, which is linearly isomorphic to the $\chi$ matrix, and is given by
\begin{equation}
    \rho_J(\cE) = (\cE\otimes\Id)\left[\proj{\Psi}\right], \label{eq:Jamiolkowski}
\end{equation}
where $\ket\Psi$ is a maximally entangled state between the system of interest and an auxiliary system of the same dimension.  However, we want metrics for error \emph{generators}, which are not processes but rather infinitesimal generators of processes.  It is very simple to compute the Jamiolkowski ``state'' for a generator $\cL$ -- just substitute $\cE \to \cL$ into the formula (Eq. \ref{eq:Jamiolkowski}) for $\rho_J$ -- but the result is not actually a density matrix.  It is approximately a \emph{difference} of density matrices, since $\cL \approx \cE-\Id$. Our metrics take account of this.

\subsection{Jamiolkowski probability}

We call the first metric \emph{Jamiolkowski probability} (or ``J-probability'' for short).  It is the total probability created by $\cL$ on the orthogonal complement to $\ket{\Psi}$ (see Fig.~\ref{fig:JamiolkowskiMetrics}):
\begin{eqnarray}
    \PJ(\cL) &=& \Tr\left[ \rho_J(\cL) \left(\Id - \proj{\Psi}\right)\right] \\
    &=& -\Tr\left[ \rho_J(\cL) \proj{\Psi} \right].
\end{eqnarray}
It is extremely simple to compute the J-probability for each elementary generator:  $\PJ = 1$ for every $S_P$ generator, and $\PJ=0$ for every other elementary generator.  If an error generator $L$ is represented by the list of rates $\{h_P, s_P, c_{P,Q}, a_{P,Q}\}$, then $\PJ(L) = \sum_P{s_P}$.

A generator's J-probability quantifies the average probability that it ``flips'' a state to some orthogonal state.  Geometrically, it measures shifts in $\rho$ that commute with $\rho$, and thus change $\rho$'s spectrum.  

``Average'' here means averaging over states; an input state that is maximally entangled with a reference is a fairly standard proxy for a random input state, and there is a very close relationship between this quantity and averages over pure states of the system alone (see, e.g. \cite{Nielsen2002-wz}).  The J-probability is \emph{not} a complete description of the error!  Consider, for example, the difference between (a) a dephasing process generated by $p S_Z$, (b) the stochastic process generated by $\frac{p}{2}(S_X+S_Y)$, and (c) the $T_1$ process generated by $\frac{p}{2}(S_X + S_Y + A_{X,Y})$.  All three have a J-probability of $p$, but the actual probability of an error is input-state dependent, and that dependence varies greatly over the three processes.

\subsection{Jamiolkowski amplitude}

We call the second metric \emph{Jamiolkowski amplitude} (or ``J-amplitude'' for short).  It is the total amplitude created by $\cL$ on the orthogonal complement to $\ket{\Psi}$ (see Fig.~\ref{fig:JamiolkowskiMetrics}):
\begin{eqnarray}
\theta_J(\cL) &=& \left\|\left(\Id-\proj{\Psi}\right)\rho_J(\cL)\ket\Psi\right\| \label{eq:JA}\\
&=& \sqrt{\braopket{\Psi}{\rho_J(\cL)^2}{\Psi} - \braopket{\Psi}{\rho_J(\cL)}{\Psi}^2}. \nonumber
\end{eqnarray}
To compute the J-amplitude for elementary generators, we recall that each generator is a sum of Choi units $X_{P,Q}$ that map $\rho \to P\rho Q$.  A Choi unit's action on $\proj{\Psi}$ is simple: each Pauli maps the maximally entangled state $\ket\Psi$ to a distinct and orthogonal maximally entangled state.  So if $P$ and $Q$ are both different from $\Id$, then $P\rho Q$ is supported entirely on the orthogonal complement to $\ket\Psi$, and does not contribute to Eq. \ref{eq:JA}.  Only Choi units where exactly one of $P$ or $Q$ is $\Id$ contribute.  These terms come in pairs, as required by the hermiticity of $\rho_J$, and Eq. \ref{eq:JA} only counts the terms where $P=\Id$ and $Q\neq\Id$.

Unsurprisingly, $\theta_J=1$ for every Hamiltonian generator, because $H_P$ creates imaginary amplitudes on the orthogonal complement to $\ket\Psi$.  For every Pauli-stochastic generator, $\theta_J = 0$, confirming that $S_P$ causes only incoherent errors.  The $A_{P,Q}$ generators contain (see Eq.~\ref{eq:APQ}) an anticommutator term, $(i/2)\left\{[P,Q],\rho\right\}$, that creates real amplitudes on the orthogonal complement whenever $P$ and $Q$ anticommute.  These directly reflect the affine shift produced by the same generators (consider, as in Fig.~\ref{fig:T1-decompose}, the way that a $T_1$ process shifts $X$ eigenstates in the $Z$ direction).  So for $A_{P,Q}$, $\theta_J=1$ if $\{P,Q\}=0$, or zero otherwise.  Some Pauli correlation generators also create error amplitude.  If $P$ and $Q$ commute, then $C_{P,Q}$ contains an anticommutator term, $-(1/2)\{\{P,Q\},\rho\}$, that creates real amplitudes on the orthogonal complement.  So $\theta_J=1$ for $C_{P,Q}$ if $[P,Q]=0$, or zero otherwise.

Computing $\theta_J$ for a linear combination of generators is not as simple as computing $\PJ$, because amplitudes can interfere.  But for generators that do not interfere, $\theta_J$ adds in quadrature.  For example, all of the Hamiltonian generators $H_P$ induce distinct amplitudes that do not interfere, so $\theta_J(\cL_\bH) = \left(\sum_P{h_P^2}\right)^{1/2} = \|\vec{h}\|$.

A generator's $J$-amplitude quantifies its ability to create ``coherent'' errors -- i.e., to induce changes in an input state $\rho$ that are orthogonal to $\rho$'s commutant and therefore do not change its spectrum.  Unitary errors are the most obvious such errors -- they \emph{never} change the spectrum of an input state, so their impact is purely coherent.  But other error generators can also produce such shifts, as seen above for the $A$ and $C$ generators.  The resulting coherent errors behave just like those induced by unitary error processes -- they can interfere constructively or destructively with coherent errors caused by Hamiltonian generators -- and are captured by $\theta_J$.  As with $\PJ$, $\theta_J$ is definitely \emph{not} a complete description of an error.  For example, understanding how two coherent errors combine requires knowing their phase and direction in addition to their $\theta_J$.

\subsection{Discussion}

We conclude this section with a brief discussion of how $\PJ$ and $\theta_J$ relate to process fidelity (which can be defined in different ways; we consider entanglement fidelity \cite{Barnum1998-sw}) and diamond norm.  Each of these metrics compresses \emph{all} errors in an error process $\cE$ into a single number, but in very different ways.  For small error processes, entanglement fidelity is almost exactly equal to $1- (\PJ + \theta_J^2)$.  A small error process's diamond norm error is less easy to approximate, because (unlike fidelity, J-probability, and J-amplitude), diamond norm is defined not by an average but by a maximization over all possible input states.  This breaks any direct connection to the Jamiolkowski state, and results in a metric that is nearly impossible to predict without numerics, because a process's worst-case behavior depends on the most fine-grained details of the error (e.g. whether different types of error both impact the same input state, or not).  Up to $O(1)$ multiplicative factors, however, the diamond norm error scales as $O(\PJ + \theta_J)$.

\section{Constructing reduced models with error generators} \label{sec:Reduced}

The preceding sections are a complete, self-contained presentation of the error generator representation.  We conclude, in this section, by outlining what we see as the most exciting \emph{application} of the error generator representation.  The most obvious use of error generators is a tool to analyze process matrices obtained from modeling, simulation, or tomography.  We have used them as such, to understand estimates derived from gate set tomography \cite{Blume-Kohout2017-kn, Nielsen2020-cu, Nielsen2020-lu} since 2017.  But error generators can also be used to construct parameterized models for gate errors that are simpler, sparser, and more efficient than process matrices.  We call these \emph{reduced models}.

An $N$-qubit process has $4^N(4^N-1)$ free parameters.  This is an unwieldy number even for $N=2$, and presents absurd data-storage and computation challenges for $N\gg2$.  As described above, that process can be faithfully represented by a list of elementary error generators' rates.  This representation of $\cL$ is perfectly equivalent to $G$, and has the same number of parameters.

But in the error generator representation, it is easy to do something that is not possible (or at least not easy) for process matrices.  We can separate the elementary generators into (1) those expected to appear and/or play a significant role in realistic noise, and (2) everything else.  The latter can be discarded, setting their rates to zero by fiat.  The subspace of generators spanned by the remaining generators defines a \emph{reduced model} for gate errors.

It is easy to construct reduced models using this framework.  We can construct relatively generic reduced models, intended to describe any process whose errors respect certain principles.  We can also construct specific customized models for specific quantum processors with known physics.  An $n$-parameter reduced model for a single gate is simply a $n$-dimensional subspace $\bM\subset\bL$.  This model's $n$ parameters are the rates (coefficients) of basis vectors (elementary generators, or linear combinations of them) that span $\bM$.  

The easiest way to construct such a subspace is by simply making a list of elementary error generators, and defining the reduced model $\bM$ as their span.  It is also sometimes useful or necessary to include specific linear combinations of elementary error generators (e.g., $S_X + S_Y + S_Z$ for a single qubit describes depolarization).  We can also construct reduced models for an entire \emph{gate set} -- a list of CPTP maps describing all the operations exposed by a processor's API in a common reference frame \cite{Nielsen2020-cu} -- by simply specifying $\bM$ for each gate.  Often, the same reduced model is used for every gate, but in other cases the physics of the system suggests that different error generators should be ``activated'' (included in $\bM$) or ``frozen'' (excluded from $\bM$) for different gates.

A very easy way to construct reduced models is to build $\bM$ from entire sectors.  For example, we have found the ``H+S'' model 
\begin{equation}
    \bM_{H+S} = \bH\oplus\bS
\end{equation}
to be useful for 1- and 2-qubit Pauli-rotation gates ($\bar{G}$ is a Pauli rotation if $\bar{G}[\rho] = e^{-i\theta P}\rho e^{i\theta P}$ for some Pauli $P$).  It can model all unitary errors (including over/under-rotations as well as ``tilt'' errors that change the gate's rotation axis), and the most common stochastic errors including depolarization and dephasing in the gate's eigenbasis (as is produced, e.g., by fluctuating over/under-rotation).  For a single qubit, this model has $2\times$ fewer parameters than a full CPTP map; for two qubits it is $8\times$ more efficient.

\begin{figure*}[t!]
\includegraphics[width=1.6\columnwidth]{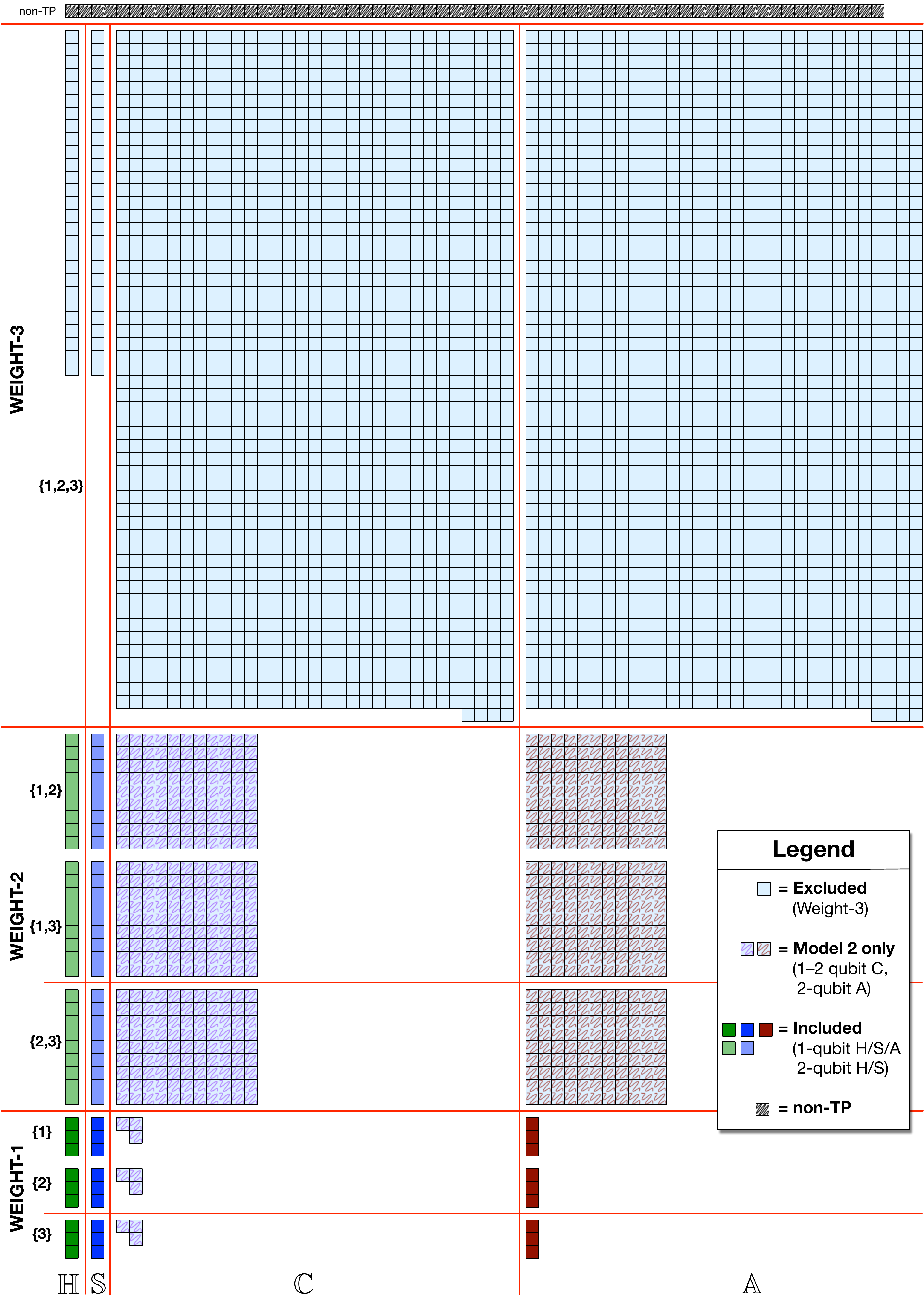}
\caption{For a generic 3-qubit error process, this figure illustrates the relative sizes of the $\bH / \bS / \bC / \bA$ sectors and subsectors of its error generator space.  Each small box represents one of the $4096$ elements in a $64\times 64$ process matrix.  64 are constrained by trace preservation, leaving 4032 free parameters.  More than $83\%$ of these ($27+27+1647+1647=3348$) are weight-3, and excluded from both the ``Weight$\leq 2$'' and ``H2+S2+A1'' models described in the main text.  Another $15\%$ (297+297+9=594) are weight-1/2 $C$ and weight-2 $A$-type generators excluded from ``H2+S2+A1'' .  The remaining 81 generators in the ``H2+S2+A1'' model constitute just $2\%$ of the full generator space, but represent most phenomena appearing in physical models of error.}
\label{fig:3Q-partition}
\end{figure*}

For $N>1$ qubit processors, generator space can be partitioned further.  This fine-grained partition becomes increasingly useful as $N$ grows.  We partition each sector ($\bH,\bS,\bC,\bA$) by the \emph{weight} and \emph{support} of its elementary generators (see Fig.~\ref{fig:2Q-Partition}c).  These are defined as follows:
\begin{enumerate}
    \item The \emph{support} of Pauli operator $P$ is the set of qubits on which it acts nontrivially.  
    \item The support of a generator $H_P$ or $S_P$ is the support of $P$.
    \item The support of a generator $C_{P,Q}$ or $A_{P,Q}$ is the union of the supports of $P$ and $Q$.
    \item The \emph{weight} of a generator is the number of qubits in its support.
\end{enumerate}
Each elementary generator can be unambiguously labeled by its support (and thus its weight).  We can then partition each sector into $N$ distinct subsectors of fixed weight $w=1\ldots N$.  We refer to these subsectors as $\bH_w$, $\bS_w$, $\bC_w$, and $\bA_w$ respectively.  If desired, we can partition each of those subsectors into $\binom{N}{w}$ subsectors of fixed support $\bQ$.

This fine-grained partition of generator space provides a great deal of flexibility to construct models that (1) respect either general locality principles or specific physical modeling assumptions, and (2) have relatively few parameters.  These models form a lattice, bookended by the ``full CPTP model'' $\bL$ and the ``target model'' $\emptyset$.  Exploring this lattice in detail is beyond the scope of this article, but here (and illustrated in Fig.~\ref{fig:3Q-partition}) are a few examples that suggest the framework's potential .
\begin{enumerate}
    \item \textbf{The 2-qubit ``H+S+A1'' model}.  This is a reduced model for 2-qubit subsystems that incorporates all the errors (including crosstalk and $T_1$ decay) predicted by most theory models.  It includes the entire $\bH$ and $\bS$ sectors on both qubits (15 parameters each), and the weight-1 $\bA_1$ subsector required to model local amplitude-damping errors on both qubits (3+3=6 parameters).  This model has 36 parameters, compared to the 240 required for a process matrix.
    \item \textbf{All weight$\leq2$ errors on $N$ qubits}.  A reasonable ansatz for a single circuit layer of non-entangling gates on an $N$-qubit processor is that no 3-body couplings exist, and therefore that the dynamics can be well-approximated by local (weight-1) and 2-qubit (weight-2) errors.  This suggests a reduced model $\bM = \bH_1\oplus\bH_2 \oplus \bS_1\oplus\bS_2 \oplus \bC_1\oplus\bC_2 \oplus \bA_1\oplus\bA_2$.  It contains 12 weight-1 error generators on each qubit, and 216 weight-2 error generators on each pair, for a total of $108N^2-96N$ parameters.  This is a nontrivial number, but for $N=10$ qubits, $\mathrm{dim}(\bM)$ is only 9840, which is tractable on modern computers.  A 10-qubit CPTP map has just over $10^{12}$ parameters, which is not tractable.
    \item \textbf{The $N$-qubit ``H2+S2+A1'' model}.  We can combine the virtues of the previous two models to get a more efficient $N$-qubit model that excludes Pauli-correlation and weight-2 active error generators.  This model is formally given by $\bM = \bH_1\oplus\bH_2 \oplus \bS_1\oplus\bS_2 \oplus \bA_1$.  It contains 9 weight-1 error generators on each qubit, and 18 weight-2 error generators on each pair, for a total of $9N^2$ parameters.  For $N=20$ qubits, it has 3249 parameters.
\end{enumerate}

These models are examples, but more importantly they are starting points.  Given any specific architecture, it's easy to point out specific errors that are likely to occur in that system, but not included in the models above.  But it is even easier to choose and add $O(1)$ elementary error generators to the model, so that it \emph{can} model that effect -- without adding very much to the model's complexity.

\section{Conclusions}

Several of the ideas appearing in this article are well-known.  Neither small-error processes \cite{Korotkov2013-bz} nor examining generators of dynamical maps \cite{Lindblad1976-qr,Bellomo2009-nr, Schirmer2010-bc, Pollock2018-ys, Gentile2020-dk} are new ideas.  But by building on these basic ideas, and using them to classify \emph{all} Markovian error generators, it becomes possible to (1) analyze experimentally reconstructed logic gates, (2) dissect their errors into distinct components with simple physical interpretations, and (3) construct simpler parameterized error models from subsets of those components.  Our development of this framework was inspired in large part by the demands of experimental tomography (e.g. \cite{Blume-Kohout2017-kn, Kim2015-ap, Dehollain2016-zt, Mavadia2018-al, Ware2018-cq, Hughes2020-wp}, but also notably by Ref. \cite{Kueng2016-vy}, which elucidated the behavior of single-qubit processes in a similar way.

We have been using these techniques extensively in our own research, beginning more than 3 years ago, and have found them to be useful and robust.  We are optimistic that physically-motivated reduced models constructed using error generators will -- finally -- make it not just possible but \emph{easy} to comprehensively describe, measure, and reconstruct the real-world dynamics of $N$-qubit systems for $N=10$, 20, and beyond.  We hope the quantum computing community will find these techniques useful too.

\begin{acknowledgments}
This work was supported by the U.S. Department of Energy, Office of Science, Office of Advanced Scientific Computing Research Quantum Testbed Program, and the Office of the Director of National Intelligence (ODNI), Intelligence Advanced Research Projects Activity (IARPA).  MPS was partially funded by LPS/ARO grant W911NF-14-C-0048 (while employed at Raytheon BBN Technologies).  RBK thanks Ivan Deutsch for pointing out Ref.~\cite{Korotkov2013-bz}.  Sandia National Laboratories is a multimission laboratory managed and operated by National Technology and Engineering Solutions of Sandia, LLC., a wholly owned subsidiary of Honeywell International, Inc., for the U.S. Department of Energy's National Nuclear Security Administration under contract DE-NA-0003525. All statements of fact, opinion or conclusions contained herein are those of the authors and should not be construed as representing the official views or policies of IARPA, the ODNI, the U.S. Department of Energy, or the U.S. Government.
\end{acknowledgments}

\bibliography{taxonomy}

\end{document}